 \documentclass[sigconf]{acmart}
\AtBeginDocument{%
  }

\copyrightyear{2026}
\acmYear{2026}
\setcopyright{cc}
\setcctype{by}
\acmConference[CHI '26]{Proceedings of the 2026 CHI Conference on Human Factors in Computing Systems}{April 13--17, 2026}{Barcelona, Spain}
\acmBooktitle{Proceedings of the 2026 CHI Conference on Human Factors in Computing Systems (CHI '26), April 13--17, 2026, Barcelona, Spain}
\acmPrice{}
\acmDOI{10.1145/3772318.3791675}
\acmISBN{979-8-4007-2278-3/2026/04}

\usepackage[dvipsnames]{xcolor}
\newcommand{\eg}{{\it e.g.,\ }}

\newcommand{\ie}{{\it i.e.,\ }}

\newcommand{\name}{{\it InkIdeator}}
\usepackage{booktabs}
\usepackage{xargs} 
\usepackage{soul}  
\usepackage{color} 
\usepackage{xspace}
\usepackage{graphicx}
\usepackage{pdfpages}
\usepackage{listings}
\usepackage{enumitem}
\usepackage{booktabs}
\usepackage{array}
\usepackage{multirow}
\usepackage{makecell}
\usepackage[normalem]{ulem}
\useunder{\uline}{\ul}{}
\usepackage{graphicx}
\usepackage{calc}
\usepackage{tablefootnote}

\newcommand{\pzh}[1]{\textcolor{black}{#1}}
\newcommand{\peng}[1]{\textcolor{black}{#1}}
\newcommand{\penguin}[1]{\textcolor{black}{#1}}
\newcommand{\shiwei}[1]{{\color{black} #1}}
\newcommand{\needupdate}[1]{\textcolor{black}{#1}}
\newcommand{\Ziyao}[1]{\textcolor{black}{#1}}
\newcommand{\wu}[1]{\textcolor{black}{#1}}
\newcommand{\rochelle}[1]{\textcolor{black}{#1}}
\newcommand{\minor}[1]{\textcolor{black}{#1}}
\definecolor{symbol}{HTML}{368FD2}
\definecolor{emotion}{HTML}{F8B642}
\definecolor{style}{HTML}{61C19E}
\definecolor{composition}{HTML}{BF77D9}

\begin{document}

\title{\pzh{InkIdeator: Supporting Chinese-Style Visual Design Ideation via AI-Infused Exploration of Chinese Paintings}}
\author{Shiwei Wu}
\email{wushw28@mail2.sysu.edu.cn}
\orcid{0009-0004-3929-455X}
\affiliation{%
  \institution{Sun Yat-sen University}
  \city{Zhuhai}
  \country{China}
}

\author{Ziyao Gao}
\email{zgao662@connect.hkust-gz.edu.cn}
\orcid{0009-0007-2424-0796}
\affiliation{%
  \institution{Hong Kong University of Science and Technology (Guangzhou)}
  \city{Guangzhou}
  \country{China}
}

\author{Zhendong He}
\email{hezhd6@mail2.sysu.edu.cn}
\orcid{0009-0007-7456-5863}
\affiliation{%
  \institution{Sun Yat-sen University}
  \city{Zhuhai}
  \country{China}
}

\author{Zongtan He}
\email{hezt7@mail2.sysu.edu.cn}
\orcid{0009-0005-7947-4464}
\affiliation{%
  \institution{Sun Yat-sen University}
  \city{Zhuhai}
  \country{China}
}

\author{Zhupeng Huang}
\email{huangzhp36@mail2.sysu.edu.cn}
\orcid{0009-0003-6455-1282}
\affiliation{%
  \institution{Sun Yat-sen University}
  \city{Zhuhai}
  \country{China}
}

\author{Xia Chen}
\email{chenx776@mail2.sysu.edu.cn}
\orcid{0009-0003-6300-8966}
\affiliation{%
  \institution{Sun Yat-sen University}
  \city{Zhuhai}
  \country{China}
}

\author{Wei Zeng}
\email{weizeng@hkust-gz.edu.cn}
\orcid{0000-0002-5600-8824}
\affiliation{%
  \institution{Hong Kong University of Science and Technology (Guangzhou)}
  \city{Guangzhou}
  \country{China}
}
\affiliation{%
  \institution{The Hong Kong University of Science and Technology}
  \city{Hong Kong SAR}
  \country{China}
}

\author{Xiaojuan Ma}
\email{mxj@cse.ust.hk}
\orcid{0000-0002-9847-7784}
\affiliation{%
  \institution{Hong Kong University of Science and Technology}
  \city{Hong Kong}
  \country{China}
}

\author{Zhenhui Peng}
\authornote{Corresponding author.}
\email{pengzhh29@mail.sysu.edu.cn}
\orcid{0000-0002-5700-3136}
\affiliation{%
  \institution{Sun Yat-sen University}
  \city{Zhuhai}
  \country{China}
}

\renewcommand{\shortauthors}{Shiwei Wu, Ziyao Gao, Zhendong He, Zongtan He, Zhupeng Huang, Xia Chen, Wei Zeng, Xiaojuan Ma, Zhenhui Peng.}

\begin{abstract}
\pzh{
Visual designers often seek inspiration from Chinese paintings when tasked with creating Chinese-style illustrations, posters, etc. Our formative study (N=10) reveals that during ideation, designers learn the cultural symbols, emotions, compositions, and styles in Chinese paintings but face challenges in searching, analyzing, and integrating these dimensions. 
This paper leverages multi-modal large models to annotate the value of each dimension in 16,315 Chinese paintings, built on which we propose InkIdeator, an ideation support system for Chinese-style visual designs. 
InkIdeator suggests cultural symbols associated with the task theme, provides dimensional keywords to help analyze Chinese paintings, and generates visual examples integrating user-selected keywords. 
Our within-subjects study (N=12) using a baseline system without extracted dimensional keywords, 
along with two \penguin{extended use cases by Chinese painters}, indicates InkIdeator's effectiveness in creative ideation support, helping users efficiently explore cultural dimensions in Chinese paintings and visualize their ideas. 
We discuss implications for supporting culture-related visual design ideation with generative AI.
}

\end{abstract}


\maketitle

\section{Introduction}
\begin{figure*}
  \centering
  \includegraphics[width=\linewidth]{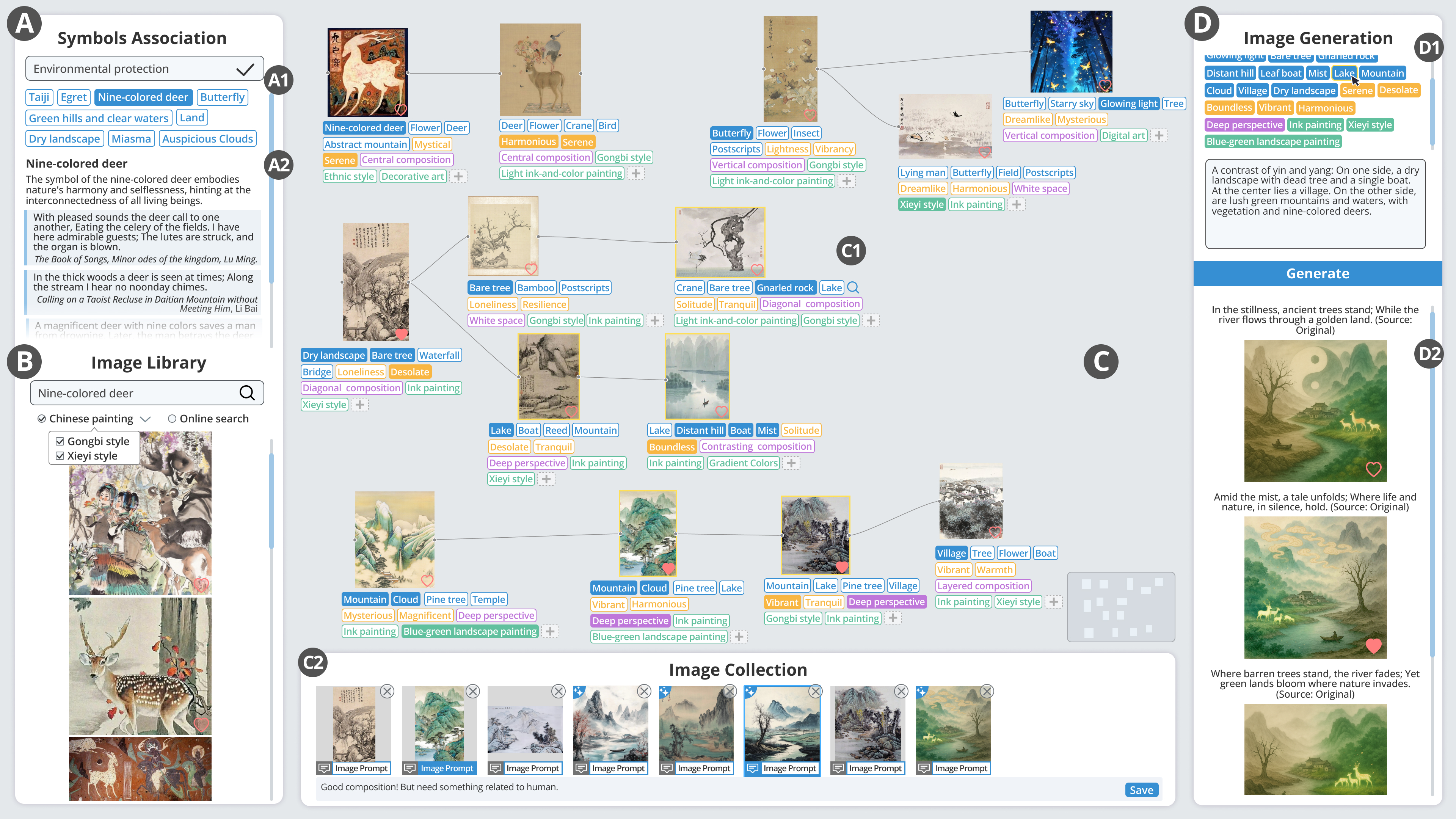}
  \caption{
  \Ziyao{
  \textbf{The interface of \textit{InkIdeator} including: }
  The Symbols Association Panel (A) recommends and explains cultural symbols based on the input design theme. The Image Library (B) allows searching examples from both our annotated Chinese painting dataset and online sources. The Moodboard Canvas (C) provides a space to arrange and analyze images by \textcolor{symbol}{cultural symbols}, \textcolor{emotion}{emotions}, \textcolor{composition}{compositions}, and \textcolor{style}{styles} to inspire association. 
  The Image Generation Panel (D) applies the explored keywords to generate sketches. 
  This interface is originally in Chinese and translated in English for presentations in this paper. 
  }
  }
  \label{fig: interface}
  \end{figure*}
\Ziyao{
Chinese-style visual design has gained increasing popularity in recent years, appearing in poster design, product packaging, book illustrations, and even game and film design~\cite{wang2025study, li2024study, wang2022implementation}. 
To create such works, designers often get inspired by Chinese painting examples, which contain cultural symbols and their aesthetic expressions ~\cite{wang2025between, li2025expressing, lyu2025cultural} that can be used in Chinese-style visual design.  
For instance, Wang Mian’s \textit{Fragrant Snow at Broken Bridge} conveys the spirit of perseverance through the depiction of a single plum blossom branch, using restrained ink tones and balanced white-space composition~\cite{WangMian_2024}.
\shiwei{In other words, Chinese paintings can serve as good visual examples that help designers learn from Chinese aesthetic expression and start ideation from previous creations rather than starting from scratch \cite{bonnardel2005towards, eckert2000sources, siangliulue2015toward}.}
}
\penguin{
This paper targets users who need to conduct Chinese-style visual design tasks and focuses on their ideation process with Chinese paintings. 
\wu{These users, regardless of whether receiving training in Chinese painting or not, can participate in Chinese-style visual design like creating posters about traditional festivals and contributing ideas about movies of Chinese tales. We note them as visual designers thereafter. }
}

\peng{
Exploring Chinese paintings for 
design ideation can be challenging in several aspects.
}
\Ziyao{
First, Chinese paintings involve a vast and diverse set of conceptual and visual elements embedded in their cultural context \cite{li2025expressing, lyu2025cultural}. 
When starting with a vague idea or task theme, as is common in early-stage design~\cite{kang2021metamap}, designers may find it difficult to decide which elements to draw upon and to locate examples that align with their intent \cite{wang2025aideation}.
}
\peng{
Second, the resources of Chinese paintings are scattered online normally with titles, authors, and basic descriptions, but without detailed interpretation about what cultural symbols are used and how they are visually integrated to convey certain spirits. 
Even though visual designers may have basic aesthetic literacy and understanding of Chinese culture, 
\penguin{many of them lack advanced training on appreciating and creating Chinese paintings}
making it inefficient in interpreting and learning the complex intertwining between cultural symbols and visual effects. 
Moreover, designers can struggle to recombine the elements \cite{choi2024creativeconnect} and convert the ideas into concrete visual representations \cite{wang2025harmonycut}. 
Researchers have developed tools (\eg MetaMap \cite{kang2021metamap}, HarmonyCut \cite{wang2025harmonycut}, Influencer \cite{chi25_influencer}) to support exploration of existing examples from multiple dimensions and generate new visual examples given textual description in the ideation process. 
Nevertheless, little effort has been spent in understanding and addressing the \penguin{ideation challenges with Chinese paintings for Chinese-style visual design}, a field that has unique domain knowledge and can boost the cultural transmission. 
}

\pzh{
To this end, we conducted a formative study with 
ten visual designers with varying experience in Chinese painting to understand the practices and challenges of ideation with Chinese paintings.
The findings indicated that designers primarily looked for the cultural symbols in the Chinese paintings, \penguin{as well as} the conveyed emotions, compositions, and overall styles like color tones and brushstrokes. 
\peng{Designers reported challenges in searching Chinese paintings with cultural symbols relevant to their design tasks, analyzing the domain-specific dimensions in the paintings, and imagining the visual effects of their ideas.}
To tackle these challenges, we first collected a dataset of Chinese paintings shared in social media and iterated the prompts to the multimodal large model \shiwei{GPT-4o mini}
with Chinese painters' feedback, resulting in $16,315$ Chinese paintings with annotated cultural symbols, emotions, compositions, and styles. 
Then, we propose \name{} (\autoref{fig: interface}), an ideation support tool \penguin{for Chinese-style visual design.}  
With \name{}, users can start the example search by getting suggested cultural symbols relevant to the theme of design tasks (\autoref{fig: interface}A), iteratively search and analyze the design dimensions of Chinese paintings in our annotated dataset (\autoref{fig: interface}B\&C), and integrate their ideas on the design dimensions into generated visual examples (\autoref{fig: interface}D). 
}

\pzh{
To evaluate \name{}, we conducted a \penguin{within-subjects study} with 12 visual designers \penguin{on given tasks,} 
comparing our tool to a baseline with the original components (\ie ChatGPT + image search on our dataset + MidJourney) in our tool but without the annotations on Chinese paintings. 
\peng{
The results indicated that while ideas from both \name{} and baseline system were comparably clear and appealing, participants reported receiving more creative support from \name{}.
\name{} did significantly better in helping  participants explore visual references in an organized way, efficiently extract interesting design elements from Chinese paintings, and efficiently turn ideas into visual representations.}
\shiwei{
\penguin{To explore generalized use cases of \name{},}
we further invited two experienced Chinese painters to ideate a Chinese painting with \name{} based on their interests. The results indicated that they actively iterated their ideas with \name{} and input independent thoughts in this ideation process, 
\penguin{demonstrating the potentials of generalizing \name{} to other culture-related creative tasks}.
}
}

\pzh{
In summary, our main contributions are as follows: 
\begin{itemize}
    \item \name{}, an ideation support tool that facilitates Chinese-style visual design ideation with Chinese paintings.
    \item \penguin{A within-subjects user study with 12 participants of varying Chinese painting expertise}
    that demonstrate the usefulness of \name{} to facilitate creative ideation of Chinese-style visual designs. 
    \item A design space \footnote{We have provided the AI-annotated design space in the supplementary materials.} regarding the cultural symbols, emotions, compositions, and styles of Chinese paintings, drawn from our formative study and AI-annotated $16,315$ examples. 
\end{itemize}
}

\section{Related Work}
\begin{table*}[]
    \centering
    \caption{
    \textbf{\Ziyao{Demographic of participants in the formative study.}} 
    }
    \label{tab:formativesusertable}
    \scalebox{0.9}{
    \begin{tabular}{cccm{13.5cm}}
        \toprule
        ID  & Gender & Age & \multicolumn{1}{c}{Chinese-style Design Experience}\\
        \midrule
        S01 & M & 86 & Famous Chinese painter (>50 years), specializing in Shanshui Chinese painting \\ \midrule
        S02 & M & 63 & College teacher, teaching Chinese painting (>20 years)  \\ \midrule
        S03 & M & 45 & College teacher, teaching Chinese painting (>20 years)  \\ \midrule
        S04 & F & 21 & Senior undergraduate students, majoring in fine arts \penguin{with courses about Chinese painting}   \\ \midrule
        S05 & M & 21 & Senior undergraduate students, majoring in fine arts \penguin{with courses about Chinese painting}  \\ \midrule
        S06 & F & 26 & Graduate student of Chinese painting, \penguin{recognized as young gongbi painter with multiple national awards (\eg Gold Award at the Cross-Strait Youth Contemporary Art and Cultural Creativity Exhibition \tablefootnote{\url{https://www.straitculture.com.cn/about.html}})}   \\ \midrule
        S07 & M & 60 & S01's apprentice with 2-year Chinese painting practice  \\ \midrule
        S08 & F & 24 & Graduate student majoring in design with experience in Chinese-style design (\eg developed a Dunhuang-themed poster in a Chinese aesthetic style) \wu{and had a course in Chinese painting for one semester}    \\ \midrule
        S09 & F & 23 & Graduate student majoring in design with experience in Chinese-style design (\eg created illustrations inspired by intangible cultural heritage) \\ \midrule
        S10 & F & 24 & Graduate student majoring in design with experience in Chinese-style design (\eg designed a poster on the theme of New Year's Animals in ink and wash style)  \\
        \bottomrule
    \end{tabular}
    }
\end{table*}

\subsection{Chinese-Style Visual Design and Its Support Tools} 
Chinese-style visual design is a creative form that integrates Chinese cultural symbols, philosophy, history, and traditional aesthetics into graphic design \cite{geng2014application},  video games \cite{miao2024expression}, movies (\eg background scenes in ``NeZha 2'' \footnote{\url{https://en.wikipedia.org/wiki/Ne_Zha_2}}), and so on.
Creators of these designs commonly draw inspiration from cultural references like poems, paintings, and folk tale. 
Among these, Chinese paintings stand out as they not only contain cultural symbols that reflect ancient Chinese views on nature, society, and philosophy \cite{feng2022ipoet}, but also have visual aesthetics that can be directly adapted to visual designs. 
For example, the Chinese painting theory of ``Woyou'' (travel in the nature) can provide theoretical support for optimizing game visual design in video games \cite{miao2024expression}. 


HCI researchers have proposed various tools to support the creation of Chinese painting \cite{hoang2024artvista, haoran2023magical, 9423340, zhang2025ink, chan2002two, ZHANG2025104330}.
For example, 
\citet{yao2024inkbrush} developed InkBrush, which provides a digital calligraphy brush and editing tools to render authentic ink strokes with attributes such as hairy edges, ink drips, and scattered dots.
\citet{haoran2023magical} introduced Magical Brush, which combines symbolic cultural factors with AI generative models to help novices create modern Chinese painting and experience cultural connotations in the process. 
Nevertheless, none of these previous works specifically focus on the ideation stage of creative activities associated with Chinese painting. 
Questions remain as how Chinese-style designers seek inspiration from Chinese painting, what challenges they encounter, and how to support them in the ideation process. 
\pzh{This paper addresses these questions via a formative study and proposed \name{}. 
}

\subsection{Example-Based Ideation Support}

Previous example-based ideation support tools commonly help designers to understand and explore the design space \cite{davis2024fashioning, Suh24Luminate, park2025leveraging}, which covers the possible ideas and solutions to a task and problem and plays an essential role in creative processes \cite{dove2016argument, heape2007design, Suh24Luminate}.
For example, \citet{kang2021metamap} developed MetaMap, which inspires visual metaphor ideation through multi-dimensional (\ie semantics, color, and shape) example-based exploration with a mindmap-like structure. 
Their user study showed that MetaMap supported diverse exploration and helped users create diverse and creative ideas \cite{kang2021metamap}.
HarmonyCut \cite{wang2025harmonycut} helped users translate abstract intentions into creative and structured ideas with the help of identified design space of paper-cutting design.
Their evaluation demonstrated that HarmonyCut effectively supported the ideation of paper-cutting designs and maintained design quality within the design space to ensure harmony between form and cultural connotation \cite{wang2025harmonycut}.
\citet{choi2024creativeconnect} proposed CreativeConnect, which helped graphic designers identify useful elements from the reference image via keywords, 
facilitated diverse recombination options based on
user-selected keywords, and showed recombinations as generated sketches paired with text descriptions. 

Overall, these ideation support tools commonly enable users to explore examples via multi-dimensional keywords of design space and an interactive mood board. 
They guide users in ideating with features aiding design space exploration, such as keyword extraction \cite{choi2024creativeconnect} and keyword-based search \cite{kang2021metamap}.
Similarly, \name{} supports visual designers in exploring the design space with a mood board.
Unlike previous work, we target Chinese-style design inspired by Chinese paintings, where the dimensions considered during the design process may be different.
To address this, we identified the design space of Chinese paintings with Chinese painters and leveraged multi-modal large models to analyze the related dimensions in the paintings. 
Our machine-annotated dataset of Chinese paintings can serve as a starting point for other Chinese-style creative activities.

\subsection{Generative AI for Creative Support}

Trained on vast datasets, Generative AIs (GenAIs) possess the ability to deeply understand diverse cultural contexts \cite{wang2025harmonycut}, analyze aesthetic features of images \cite{choi2024creativeconnect}, 
interpret fuzzy user design intention \cite{hou2024c2ideas}, and generate artistic images that meet various needs. 
Together, GenAIs can accelerate
creative tasks with abundant sources of inspiration \cite{boucher2024resistance, ko2023large, wang2025aideation}, particularly during the early ideation stages of design \cite{wang2025aideation}. 
For example, AIdeation \cite{wang2025aideation} supports early ideation by brainstorming design concepts with flexible searching and recombination of reference images.
GenQuery \cite{son2024genquery} helps user precisely express their visual references search intents with the generated image.
PopBlends \cite{wang2023popblends} automatically generates connecting concepts and scenes based on user input to facilitate concept association. 
GenAIs also can be used to extract key elements within images to facilitate images search \cite{park2025leveraging} and concept recombination \cite{choi2024creativeconnect}.

However, simply using GenAIs in the design process may lead to certain issues. 
Its ``end-to-end'' nature contrasts with the iterative, reflective practices of human creativity \cite{zhang2024confrontation}, potentially leading to limited examples explored and design fixation.
Although the generated outputs may appear satisfactory, the limited and potentially biased training data
can result in outputs that reinforce stereotypes (\eg generating common cultural symbols).
It is important to stimulate the designer's active role in the creative process, where exploration and reflection play a crucial part.
Similar to previous work, we also use GenAIs to associate concepts, extract concepts, and generate visual ideas. 
Differently, our work targets Chinese-style visual design ideations with Chinese paintings and proposes a system named \name{} that integrates GenAIs into this culture-specific ideation process. 

\section{Formative study}
\pzh{We conducted a formative study with ten participants to understand how they learn from Chinese paintings for Chinese-style visual designs, \wu{their perceptions towards GenAIs}, and challenges in the ideation process.}
\subsection{Participants and Procedure}
\penguin{
We included participants who major in design or art or are trained for Chinese paintings for two reasons. 
First, they are \name{}'s target users who need to ideate Chinese-style visual design, \eg designers would create posters about traditional Chinese festivals, painters would need to contribute their original artworks and opinions to Chinese-style illustrations of books and backgrounds of movies \wu{(\eg Black Myth: Wukong \footnote{\url{https://en.wikipedia.org/wiki/Black_Myth:_Wukong}}, Ne Zha 2 \footnote{\url{https://baike.baidu.com/item/\%E7\%81\%B5\%E7\%8F\%A0\%E6\%95\%96\%E4\%B8\%99/66812671\#reference-2}. The website is in Chinese. It is recommended to use a browser plugin to translate it.}}). 
\wu{Second, they have various levels of training and experiences in Chinese paintings, \ie from no knowledge to some knowledge to expert, which helps us gain a comprehensive understanding of their ideation practices and compile domain knowledge.}}
\pzh{
In total, we recruited ten participants via snowballing starting from S01 and \peng{S08} (\autoref{tab:formativesusertable}). 
These participants include one famous Chinese painter (S01), two college Chinese painting teachers (S02, S03), two senior undergraduate students majoring in Fine Arts (S04, S05), one graduate student specializing in Chinese painting (S06), S01's apprentice with 2-year Chinese painting practice (S07), and three designers (S08, S09, S10) with Chinese-style visual design experience.
}
\penguin{They had varying expertise of Chinese paintings (\autoref{tab:formativesusertable}): S01-03 are experts, S04-07 had painting experiences, and S08 took related courses while S09-10 did not. 
}

\pzh{
We conducted one-hour semi-structured interviews with each participant via a video-conferencing app.
The interviews asked
what and how they learn from Chinese paintings for ideating Chinese-style visual design, followed by 
the challenges they encountered, \penguin{and their attitudes toward GenAI tools (\eg Midjourney).}
During the interviews, we actively asked participants to contextualize their responses with examples, \eg in recent design tasks. 
}
\subsection{What Designers Learn from Chinese Paintings for Visual Design Ideation}\label{sec:four_dimensions}
Participants consistently highlighted that during their ideation process, they need to explore a large amount of Chinese paintings.
They usually considered the following dimensions in Chinese paintings. 

\textbf{Cultural symbols.} The concept of the Chinese cultural symbol is a unification of ``meaning'' (subjective feelings, spirits and verve) and ``object'' (objective beings) \cite{haoran2023magical}.
It is the soul of Chinese paintings, carrying with it a wealth of history, culture, and philosophy.
Participants noted that they tend to think about cultural symbols first when developing their ideas.
S04 mentioned, ``\textit{I always refer to others' works on similar themes to see what cultural symbols they include.}''
S09 also noted, ``\textit{Design emphasis on conceptual expression, the cultural meaning behind the symbol is crucial to the conceptual construction of our works.}''

\textbf{Emotion.} Emotion refers to the spiritual impact conveyed through the interplay of cultural symbols and the visual effects of the painting.
 S03 said, ``\textit{Chinese painting emphasizes the expression of emotions.}''.
 S02 also noted, 
``\textit{Painting is a process of expression that reflects the creator's inner vitality, which is closely connected to their life experiences. We are often deeply moved and inspired by the emotions conveyed through the artwork.}''

\textbf{Composition.} Composition is the arrangement of cultural symbols within a painting.
S01 shared, ``\textit{The composition of a painting can convey a sense of rhythm, where the variation in the arrangement of objects can evoke different feelings.}''
S10 mentioned, ``\textit{I would draw inspiration from compositions of elements to create a harmonious and storytelling scene.}''

\textbf{Style.} Style refers to the overall visual effect of the painting, shaped by cultural symbols and the combined effects of visual concepts (\ie color tone, brushstroke, composition).
S05 said, ``\textit{When I see outstanding paintings, I feel inspired and wonder how the artist used their brush and ink, how the colors were combined.}''
For designers, the unique aesthetics of Chinese paintings often serve as inspiration during the early ideation process. They focus more on the overall visual impact created by the interplay of brushwork and colors, rather than the detailed techniques emphasized by painters.
\wu{\subsection{Perceptions towards GenAI Tools}} \label{sec: AI_percetion}
\penguin{
Overall, participants viewed current GenAI tools as promising but with concerns on quality and authorship.
They were impressed by the speed, quality of the generated images and inspiration they provided. 
S08 noted, ``\textit{When I am stuck in my ideation process, I ask ChatGPT for suggestions.
}''
However, they reported some generated images did not adhere to the traditional style of Chinese paintings.
S01 observed, 
``\textit{The color scheme of the generated image is too vivid, in Chinese painting, we usually express subtlety}''.
They were also concerned about the authorship of AI-generated images.
S02 noted, ``\textit{Paintings move us because they reflect the emotions and thoughts that grow within the artist's hearts. From my perspective, while AI generates the images, the final outcome would not be possible without human input.}''
}
\\
\subsection{\pzh{Challenges and Design Goals}}
\label{sec:design_goals}

\begin{figure}[]
    \centering
    \includegraphics[width=\columnwidth]{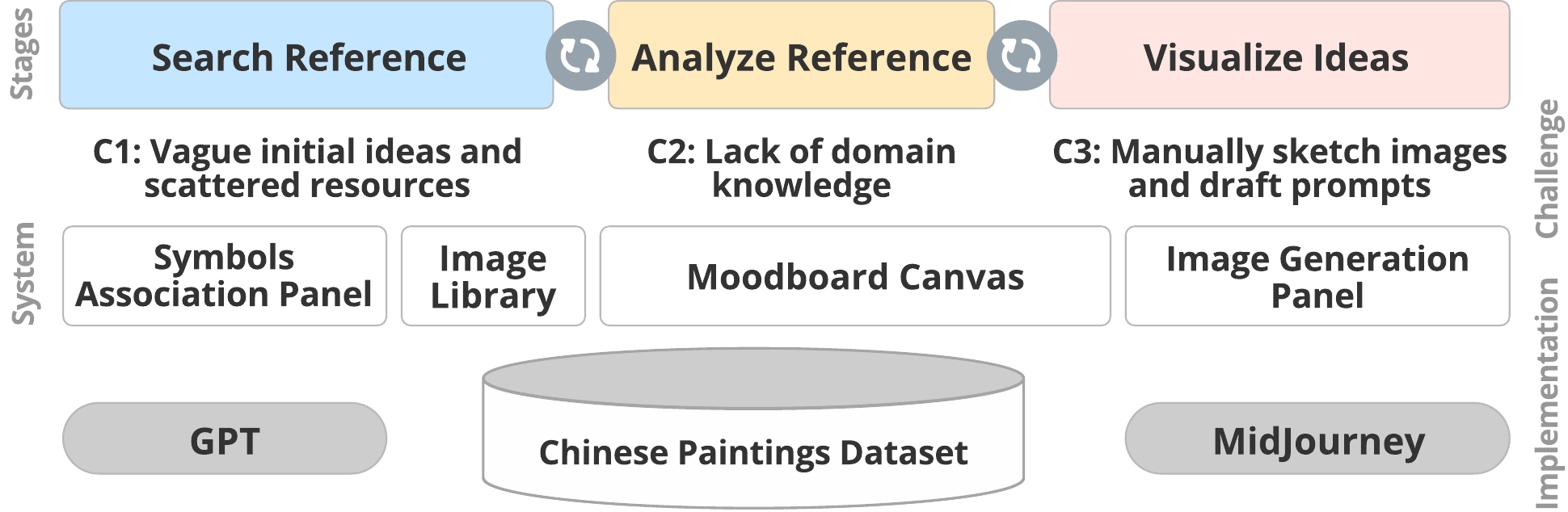}
    \caption{
    \Ziyao{
    \textbf{System overview of \textit{InkIdeator}.} 
    The system supports three iterative stages of Chinese-style visual design ideation—Search Reference, Analyze Reference, and Visualize Ideas—each addressing key challenges. 
    This is achieved through core components, including the Symbol Association Panel, Image Library, Moodboard Canvas, and Image Generation Panel, connected to a Chinese Painting Dataset and powered by GPT and MidJourney for content generation.
    }
    }
    \label{fig:stages}
\end{figure}

\pzh{Participants reported that their ideations with Chinese paintings usually consist of three iterative and challenging stages, as illustrated in \autoref{fig:stages} and detailed below.} 
\subsubsection{Designers
would like to search for Chinese paintings related to the task but found it inefficient due to vague initial ideas and scattered resources.}
\pzh{
Designers usually start ideation with a design task or a vague idea (\eg a Chinese-style poster with the theme ``longing for
one's hometown''), either given by stakeholders or self-motivated. 
To seek inspirations from Chinese paintings, they first need to find the ones that are related to their tasks. 
However, they often struggled to identify relevant cultural symbols, making it difficult for them to define the target of searching for Chinese paintings. 
The current solution is often to ``look more'', as S01 mentioned, ``\textit{The common approach is to look at others' works, take photos, and record with a camera.}''
S05 added, ``\textit{One of the challenges in the creative process is not knowing what to draw, or how to represent the theme I want to express with suitable elements.}'' 
Besides, current Chinese painting resources are also scattered online and lack systematic categorization. 
As a result, designers have to rely on searching by specific artwork titles or tags like artist names and broad styles (\eg gongbi or xieyi) in either general-purposes platforms (\eg RedNote \footnote{\url{https://www.xiaohongshu.com/explore}}) or websites specialized for Chinese paintings (\eg Zhonghua Zhenbao Museum \footnote{\url{https://www.ltfc.net/}}).
This makes the example search process inefficient, 
as S03 noted, ``\textit{Discovering a relevant painting often happens by chance when I am exploring a large number of works.}'' 
To this end, the first design goal (DG) of \name{} is: 
}


\pzh{\textbf{DG1}: Support efficient search of Chinese paintings by suggesting cultural symbols related to the task theme and compiling a structured dataset of Chinese paintings.}

\subsubsection{Designers would like to analyze the domain-specific dimensions of Chinese paintings but
often found it difficult due to lack of domain knowledge.}
When finding a relevant Chinese painting, designers usually carefully analyze it to learn the elements that can inspire them, such as cultural symbols, emotions, compositions, and styles described in \autoref{sec:four_dimensions}. 
However, many designers, especially the ones \penguin{with little or no expertise in Chinese paintings,}
reported difficulty in analyzing the Chinese paintings. 
S09 (F, 23) mentioned, ``\textit{When I view Chinese paintings, I often think about the meaning behind the symbols and the emotions conveyed, which can be quite challenging.}'' 
The limited domain knowledge 
could hinder designers' ability to come up with new ideas and expand their design space.
Therefore, we propose:

\pzh{\textbf{DG2}: Support analyses of Chinese paintings regarding cultural symbols, emotions, compositions, and styles.}
\begin{figure*}[]
    \centering
    \includegraphics[width=1.95\columnwidth]{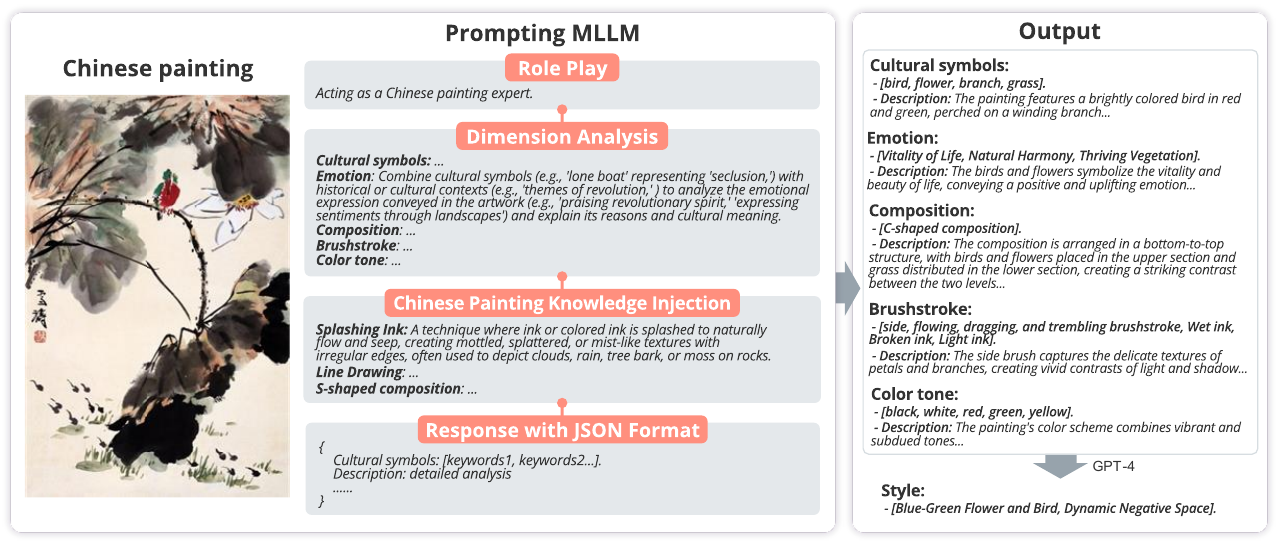}
    \caption{
    \Ziyao{
    \textbf{Prompt template and output example of analyzing Chinese paintings using the multimodal large language model (MLLM).}
    }
    The left displays the Chinese painting being analyzed, the middle panel shows the prompt structure (\ie Role Play, Dimension Analysis, Chinese Painting Knowledge Injection, and Response Format) with example, and the right panel presents the structured output analyzed by MLLM.}
    \label{fig:prompt}
\end{figure*}

\subsubsection{Designers
would like to visualize their ideas, but it is time-consuming to manually sketch them and non-trivial to draft prompts to text-to-image generation models.} 
\pzh{During the iterative ideation process, designers have ideas updated from time to time. 
For example, after they determine the usage of several symbols in the visual designs, they may have multiple ideas of applying specific composition, brushwork, and colors to convey the intended scene or emotion. 
S07 shared, \textit{``When trying to convey a particular mood or theme, there are many ways to do it, with various compositions, brushwork, and color choices. 
Different decisions can result in unexpected audiences' perceptions''}.
Traditionally, this often involves experimenting through multiple small sketches to visualize potential ideas, but this process can be time-consuming and struggling. 
S03 said, ``\textit{It's hard for me to imagine the visual effect of colors I have never seen before, leading to
fixed ways of thinking.}'' 
\peng{Prompting text-to-image generation models can be a potential solution, as already tried by S01, S04-10}.
However, drafting an effective prompt that expresses designers' ideas is non-trivial.
Therefore, we have:
}

%
\pzh{
\textbf{DG3:} Support efficient idea visualization by generating prompts to text-to-image generative models that integrate the dimensional keywords in designers' ideas. 
}
\section{Mining the Design Space in Chinese Paintings}
\label{sec: dataset}
\pzh{
The design goals center a need for a structured dataset (DG1) with annotated keywords of cultural symbols, emotions, compositions, and styles in Chinese paintings (DG2, DG3). 
This section presents how we mine the design space in Chinese paintings, yielding an annotated dataset that drives the implementation of \name{}. 
}

\subsection{Data Collection}

We crawled publicly available images posted by 17 bloggers \penguin{(\autoref{sec: blogger})}
who primarily share content related to Chinese paintings.
\penguin{To identify these bloggers, we used search keywords (\eg ``Chinese paintings'', ``Landscape paintings'') to identify relevant posts, visited the posters' homepages, and checked if they predominantly featured images of Chinese paintings.}
The reason for choosing RedNote is that participants in our formative study mentioned finding Chinese paintings on this platform, indicating that it contains reference works that help their ideation.
We filtered images using the post tags ``Chinese painting'', ``Gongbi'' and ``Xieyi'', 
resulting in a total of $16,315$ Chinese painting images.
\subsection{Data Annotation}\label{sec:data_annotation}
\begin{table*}[]
\caption{\textbf{Design space derived from the dataset with detailed cultural symbols and emotion clustering.}
Count represents the number of concepts within each category. Painting Number represents the total number of Chinese paintings that include the concepts associated of the category.}
\label{tab:designspace}
\scalebox{0.95}{
\begin{tabular}{@{}llccl@{}}
\toprule
{\color[HTML]{000000} Dimension}                                                                    & {\color[HTML]{000000} Category}                                   & \multicolumn{1}{l}{{\color[HTML]{000000} Count}} & \multicolumn{1}{l}{{\color[HTML]{000000} Painting Number}} & {\color[HTML]{000000} Example}                                                                                         \\ \hline
{\color[HTML]{000000} }                                                                             & {\color[HTML]{000000} }                                           & {\color[HTML]{000000} }                          & {\color[HTML]{000000} }                                    
& 
{\color[HTML]{000000} Distant mountains  (\autoref{fig:paintings}a, m)} \\
{\color[HTML]{000000} }  & \multirow{-2}{*}{{\color[HTML]{000000} Natural Landscape}}        & \multirow{-2}{*}{{\color[HTML]{000000} 190}}     & \multirow{-2}{*}{{\color[HTML]{000000} 3390}}              & {\color[HTML]{000000} Streams  (\autoref{fig:paintings}a), Waterfall (\autoref{fig:paintings}m)}                                           \\
{\color[HTML]{000000} }                                                                             & {\color[HTML]{000000} }                                           & {\color[HTML]{000000} }                          & {\color[HTML]{000000} }                                    & {\color[HTML]{000000} Irises  (\autoref{fig:paintings}f), Pine tree (\autoref{fig:paintings}d)}                                            \\
{\color[HTML]{000000} }                                                                             & \multirow{-2}{*}{{\color[HTML]{000000} Plants}}                   & \multirow{-2}{*}{{\color[HTML]{000000} 371}}     & \multirow{-2}{*}{{\color[HTML]{000000} 11877}}             & {\color[HTML]{000000} Lotus flower and lotus leaves  (\autoref{fig:paintings}c)}                                      \\
{\color[HTML]{000000} }                                                                             & {\color[HTML]{000000} }                                           & {\color[HTML]{000000} }                          & {\color[HTML]{000000} }                                    & {\color[HTML]{000000} Arched bridges  (\autoref{fig:paintings}e)}                                    \\
{\color[HTML]{000000} }                                                                             & \multirow{-2}{*}{{\color[HTML]{000000} Human and Life}}           & \multirow{-2}{*}{{\color[HTML]{000000} 693}}     & \multirow{-2}{*}{{\color[HTML]{000000} 7405}}              & {\color[HTML]{000000} Farmers (\autoref{fig:paintings}h), Pavilion (\autoref{fig:paintings}d)}                                            \\
{\color[HTML]{000000} }                                                                             & {\color[HTML]{000000} }                                           & {\color[HTML]{000000} }                          & {\color[HTML]{000000} }                                    & {\color[HTML]{000000} Plowing ox (\autoref{fig:paintings}h), Bird (\autoref{fig:paintings}g)}                                         \\
\multirow{-8}{*}{{\color[HTML]{000000} \begin{tabular}[c]{@{}l@{}}Cultural \\ Symbol\end{tabular}}} & \multirow{-2}{*}{{\color[HTML]{000000} Animals}}                  & \multirow{-2}{*}{{\color[HTML]{000000} 211}}     & \multirow{-2}{*}{{\color[HTML]{000000} 6833}}              & {\color[HTML]{000000} Goldfish (\autoref{fig:paintings}c), Butterfly (\autoref{fig:paintings}f)}                                           \\ \hline
{\color[HTML]{000000} }                                                                             & {\color[HTML]{000000} }                                           & {\color[HTML]{000000} }                          & {\color[HTML]{000000} }                                    & {\color[HTML]{000000} Vitality  (\autoref{fig:paintings}g)}                                          \\
{\color[HTML]{000000} }                                                                             & \multirow{-2}{*}{{\color[HTML]{000000} Praise for Nature}}        & \multirow{-2}{*}{{\color[HTML]{000000} 757}}     & \multirow{-2}{*}{{\color[HTML]{000000} 8141}}              & {\color[HTML]{000000} Natural harmony  (\autoref{fig:paintings}j)}                                   \\
{\color[HTML]{000000} }                                                                             & {\color[HTML]{000000} }                                           & {\color[HTML]{000000} }                          & {\color[HTML]{000000} }                                    & {\color[HTML]{000000} Diligence  (\autoref{fig:paintings}h), Resilience (\autoref{fig:paintings}b)}                                         \\
{\color[HTML]{000000} }                                                                             & \multirow{-2}{*}{{\color[HTML]{000000} Soul of Human Life}}       & \multirow{-2}{*}{{\color[HTML]{000000} 1186}}    & \multirow{-2}{*}{{\color[HTML]{000000} 11843}}             & {\color[HTML]{000000} Beautiful and Elegant (\autoref{fig:paintings}j)}                                         \\
{\color[HTML]{000000} }                                                                             & {\color[HTML]{000000} }                                           & {\color[HTML]{000000} }                          & {\color[HTML]{000000} }                                    & {\color[HTML]{000000} Seclusion (\autoref{fig:paintings}l), Serenity (\autoref{fig:paintings}e)}                                          \\
{\color[HTML]{000000} }                                                                             & \multirow{-2}{*}{{\color[HTML]{000000} Peaceful Atmosphere}}      & \multirow{-2}{*}{{\color[HTML]{000000} 708}}     & \multirow{-2}{*}{{\color[HTML]{000000} 4284}}              & {\color[HTML]{000000} Tranquility  (\autoref{fig:paintings}d)}                                       \\
{\color[HTML]{000000} }                                                                             & {\color[HTML]{000000} }                                           & {\color[HTML]{000000} }                          & {\color[HTML]{000000} }                                    & {\color[HTML]{000000} A sense of life (\autoref{fig:paintings}e)}                                           \\
{\color[HTML]{000000} }                                                                             & \multirow{-2}{*}{{\color[HTML]{000000} Understanding Life}}       & \multirow{-2}{*}{{\color[HTML]{000000} 1387}}    & \multirow{-2}{*}{{\color[HTML]{000000} 8387}}              & {\color[HTML]{000000} Freedom  (\autoref{fig:paintings}f), Reunion (\autoref{fig:paintings}k)}                                            \\
{\color[HTML]{000000} }                                                                             & {\color[HTML]{000000} }                                           & {\color[HTML]{000000} }                          & {\color[HTML]{000000} }                                    & {\color[HTML]{000000} The grandeur of nature and}                  \\
\multirow{-10}{*}{{\color[HTML]{000000} Emotion}}                                                   & \multirow{-2}{*}{{\color[HTML]{000000} Philosophical Reflection}} & \multirow{-2}{*}{{\color[HTML]{000000} 865}}     & \multirow{-2}{*}{{\color[HTML]{000000} 4832}}              & {\color[HTML]{000000} the insignificance of humanity  (\autoref{fig:paintings}m)} \\ \hline
{\color[HTML]{000000} }                                                                             & {\color[HTML]{000000} }                                           & {\color[HTML]{000000} }                          & {\color[HTML]{000000} }                                    & {\color[HTML]{000000} Gongbi flower-and-bird  (\autoref{fig:paintings}j)}                            \\
\multirow{-2}{*}{{\color[HTML]{000000} Style}}                                                      & \multirow{-2}{*}{{\color[HTML]{000000} -}}                        & \multirow{-2}{*}{{\color[HTML]{000000} 4162}}    & \multirow{-2}{*}{{\color[HTML]{000000} -}}                 & {\color[HTML]{000000} Blue-green landscape  (\autoref{fig:paintings}a)}                              \\ \hline
{\color[HTML]{000000} }                                                                             & {\color[HTML]{000000} }                                           & {\color[HTML]{000000} }                          & {\color[HTML]{000000} }                                    & {\color[HTML]{000000} Triangular composition (\autoref{fig:paintings}c)}                             \\
\multirow{-2}{*}{{\color[HTML]{000000} Composition}}                                                & \multirow{-2}{*}{{\color[HTML]{000000} -}}                        & \multirow{-2}{*}{{\color[HTML]{000000} 145}}     & \multirow{-2}{*}{{\color[HTML]{000000} -}}                 & {\color[HTML]{000000} Central composition  (\autoref{fig:paintings}j)}                     \\ \hline
{\color[HTML]{000000} }                                                                             & {\color[HTML]{000000} }                                           & {\color[HTML]{000000} }                          & {\color[HTML]{000000} }                                    & {\color[HTML]{000000} Thick ink (\autoref{fig:paintings}d)}                                          \\
\multirow{-2}{*}{{\color[HTML]{000000} Brushstroke}}                                                & \multirow{-2}{*}{{\color[HTML]{000000} -}}                        & \multirow{-2}{*}{{\color[HTML]{000000} 177}}     & \multirow{-2}{*}{{\color[HTML]{000000} -}}                 & {\color[HTML]{000000} Fine line drawing  (\autoref{fig:paintings}i)}                                 \\ \hline
{\color[HTML]{000000} }                                                                             & {\color[HTML]{000000} }                                           & {\color[HTML]{000000} }                          & {\color[HTML]{000000} }                                    & {\color[HTML]{000000} Malachite green (\autoref{fig:paintings}a)}                                    \\
\multirow{-2}{*}{{\color[HTML]{000000} Color Tone}}                                                 & \multirow{-2}{*}{{\color[HTML]{000000} -}}                        & \multirow{-2}{*}{{\color[HTML]{000000} 816}}     & \multirow{-2}{*}{{\color[HTML]{000000} -}}                 & {\color[HTML]{000000} Peach pink (\autoref{fig:paintings}h), Black ink (\autoref{fig:paintings}e)}                                         \\  \bottomrule

\end{tabular}
}
\end{table*}
\begin{figure*}[]
    \centering
    \includegraphics[width=1.9\columnwidth]{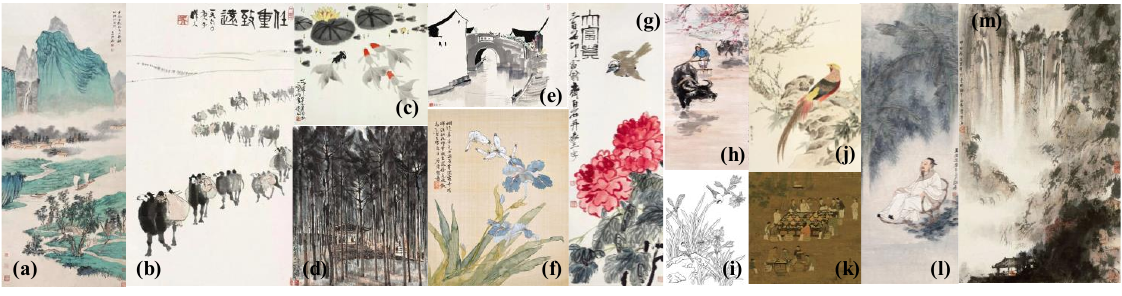}
    \caption{\wu{\textbf{Chinese painting examples} that incorporate diverse cultural symbols, express different emotions, and reflect a range of artistic techniques (\penguin{\autoref{tab:designspace}}).}}
    \label{fig:paintings}
\end{figure*}




\subsubsection{Classify the type of paintings.}
\pzh{Two authors first manually annotated the two basic types of Chinese paintings, \ie \textit{Gongbi} and \textit{Xieyi}, each with 500 images.}
We then used the MambaVision \cite{hatamizadeh2024mambavision} model to extract features from the images and split the dataset into a 7:3 ratio for training and validation sets.
We trained a classifier that includes two fully connected layers and a dropout layer to determine the type of Chinese paintings. 
Ultimately, our model achieved an accuracy of 88.0\% on the validation set.
Subsequently, we used the model to classify the remaining images, ultimately obtaining $11,645$ \textit{Xieyi} paintings and $4,670$ \textit{Gongbi} paintings.

\subsubsection{Prompt engineering.}
We leveraged GPT-4o mini, a multi-modal large language model (MLLM), to extract labels relevant to dimensions of design space from Chinese paintings.
\shiwei{Since we used the API in January 2025 without explicitly granting authorization, according to OpenAI's Consumer Privacy commitment \footnote{\url{https://openai.com/consumer-privacy/}}, the data we uploaded would not be used to train its models.}
We implemented a detailed prompting strategy following established prompt engineering principles that include four key components (\autoref{fig:prompt}): \textit{Role Play}, \textit{Dimension Analysis}, \textit{Chinese Painting Knowledge Injection} and \textit{Response with JSON Format}. 
\pzh{
Specifically, we prompted the model to act as a Chinese painting expert and first analyzed the Chinese painting from the dimensions of cultural symbols, emotion, brushstroke, and color with detailed task instructions and response format. 
As the style of Chinese paintings was shaped by cultural symbols and the combined effects of
visual concepts (\ie color tone, brushstroke, composition), we then prompt the LLM to generate style-related keywords given the analytic results on other dimensions from last step. 
}
We injected domain knowledge about Chinese painting from related literature and book \wu{\cite{Bi2013Principles}} with the definition and detailed explanations (\ie typical visual characteristics, common usage scenarios) of key concepts.

In each iteration, we evaluated the generated analyses against feedback from Chinese painters \peng{(\ie S03, S05)} from our formative study and refined our prompts accordingly. 
\penguin{
For example, we initially prompted the MLLM to analyze Chinese paintings cross five dimensions with their definitions. 
}
\penguin{
However, S05's feedback highlighted that the results lacked standard concepts commonly used in Chinese painting. Following his advice, we incorporated these concepts \penguin{(\eg S-shaped composition and layered shading)} into the prompt. 
While S05 praised the improvements, he noted that some techniques were misused, such as applying traditional xieyi methods (\eg splashing ink and breaking ink) to gongbi paintings. 
We further refined the prompt by adding ``commonly used in xieyi'' to corresponding example techniques. 
}

\subsubsection{Qualitative evaluation.}
We invited \shiwei{S05 and another two Chinese painting students (a fourth-year student and a PhD student, \wu{denoted as S11 and S12})}
to evaluate the MLLM-annotated performance. 
\penguin{Given that manually analyzing all dimensions of Chinese paintings is time-consuming (about 20 minutes per painting reported by our annotators), to cover as many Chinese paintings as possible, each annotator was assigned different paintings with different focuses.
Specifically, S05 and S11 examined all dimensions except for style of six and nine paintings, respectively. 
S12 examined 29 paintings, focusing on their descriptions of composition and brushstrokes. 
}
In total, they analyzed 44 Chinese paintings, among which the cultural symbols, color tone, and emotion of 15 paintings were analyzed, while the composition and brushstrokes of all 44 paintings were examined. 

For cultural symbols, 39 concepts were correctly annotated across 15 paintings, with 4 missed and 5 incorrectly labeled.
For color tone, 39 concepts were correctly annotated, 5 missed, and 1 incorrectly labeled.
For composition, 12 paintings were correctly annotated, while 25 had errors. Experts noted that 7 of these had atypical compositions, making evaluation difficult.
Regarding brushstrokes, it is challenging to comprehensively identify all the techniques used.  Experts assessed whether annotations were mostly correct: 21 paintings were largely accurate, while 22 had clear errors, such as missing or mislabeling techniques.
An expert commented on a painting, ``\textit{The brushwork is highly versatile, with a single stroke incorporating multiple techniques, making it impossible to define precisely.}''
For emotions, due to their subjective nature, it is difficult to define accuracy, and experts believed that some interpretations were also acceptable. 
\subsection{Design Space}
\penguin{
This subsection presents the design space (\autoref{tab:designspace}) of Chinese paintings mined from our dataset, which is applied to searching and analyzing paintings in \name{}. 
}
We first preprocessed the cultural symbols and emotional concepts by removing concepts that were not specific objects or emotions (\eg ``cultural connotation'', ``background''). 
Emotion concepts were simplified by eliminating unnecessary embellishments (\eg ``It expresses'', ``It represents'').
We also split compound concepts connected by ``and'' into independent parts (\eg ``vitality and hope'', ``farmer and ox'').
After pre-processing, we had $1265$ different concepts of cultural symbols and $4903$ different concepts of emotion.
To provide a more comprehensive view on the used symbols and conveyed emotions in Chinese paintings, we use the ``bert-base-chinese'' model \footnote{\url{https://huggingface.co/google-bert/bert-base-chinese/tree/main}} to encode the concept and used the k-means algorithm to cluster them.
Specifically, we used the elbow method to determine a reasonable range for the number of clusters, then tested clustering within this range. 
Manual evaluation was performed to check consistency within clusters and overlap between them. 
Similar clusters were merged, while those with internal differences were further divided into smaller clusters.
Based on these steps, we manually adjusted the cluster assignments to ensure that each final cluster is inherently meaningful. 

\pzh{
Overall, most paintings ($N=11877$) in our dataset contains symbols about plants (\eg irises, lotus leaves), followed by those about human and life that include people in different states or roles
and daily objects ($N=7405$, \eg arched bridges, farmers), animals ($N=6833$, \eg plowing ox, goldfish), and natural landscape ($N=3390$, \eg mountains, streams). 
Most paintings ($N=11843$) convey emotions about soul of human life that praises for human qualities (\eg diligence, resilience), followed by emotions about understanding life ($N=8387$, emotions tied to specific life experience, \eg freedom, reunion) and praising for nature ($N=8141$, including praise for different seasons and awe for extraordinary natural scenes, \eg vitality, natural harmony). 
Many paintings in our dataset also convey philosophical reflection on life ($N=4832$, \eg grandeur of nature and insignificance of humanity) and demonstrate a peaceful atmosphere with senses of tranquility and serenity ($N=4284$, \eg seclusion, tranquility). 
For composition, style, brushstroke, and color, we did not cluster them because they are either specific terms (\eg ``thick ink'', ``outline drawing'') or combinations derived from such terms (\eg  ``Light Ochre and Freehand Painting'' is a style that combines soft, delicate coloring with free-flowing brushwork.), which are difficult to group into meaningful categories. 
Nevertheless, our dataset have rich keywords describing the styles ($N=4162$), composition ($N=145$), brushstroke ($N=177$), and color tone ($N=816$) in Chinese paintings.}
\begin{figure*}[h]
    \centering
    \includegraphics[width=2\columnwidth]{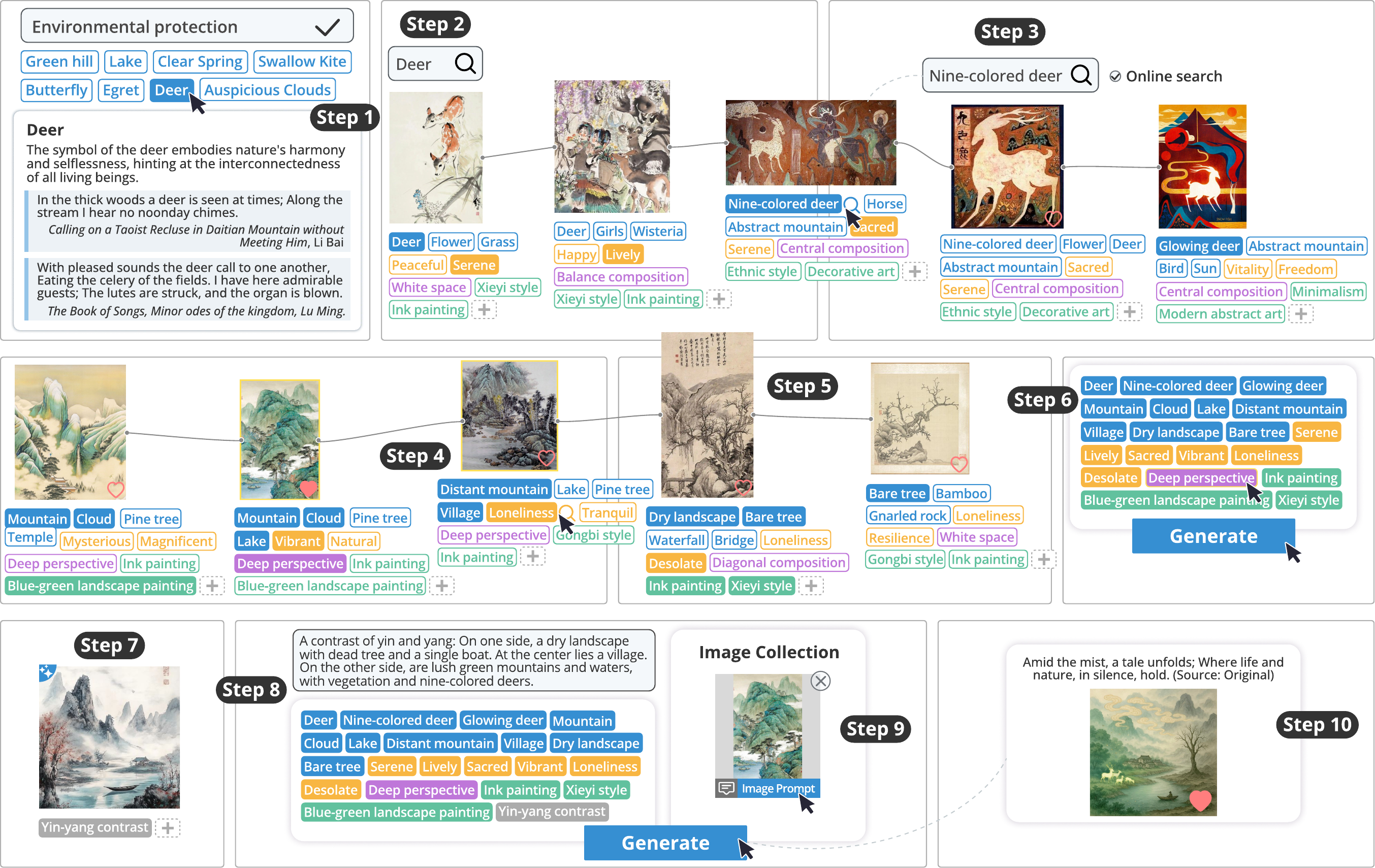}
    \caption{
    \Ziyao{
    \textbf{Interaction with \textit{InkIdeator} in user scenario. }
    Steps 1–3: Diverge on a symbol under the environmental protection theme and ground it in the cultural context to launch ideation, then search example images to refine it. 
    Steps 4–5: Analyze references from multiple perspectives of symbol, emotion, composition, and style, to promote the remaining search. 
    Steps 6–10: Apply explored keywords to iteratively generate sketches. 
    }
    }
    \label{fig:UserScenario}
\end{figure*}
\section{Design and Implementation of InkIdeator}
\pzh{Based on the derived design goals from formative study and powered by our annotated dataset, we develop \name{} to
facilitates Chinese-style visual design ideation with Chinese paintings.}
\pzh{\name{} has \peng{four} panels,
\ie a Symbol Association Panel (\autoref{fig: interface}A) for suggesting cultural symbols related to the task theme (DG1), an Image Library (\autoref{fig: interface}B) that allows keyword-based example search (DG1), a MoodBoard Panel (\autoref{fig: interface}C) that helps analyze Chinese paintings and continue searching with dimensional keywords (DG1, DG2), and an Image Generation Panel (\autoref{fig: interface}D) that visualizes user ideas given dimensional keywords or user-specified prompts (DG3). 
}

\subsection{Interface}
\name{} is implemented as a web-based application using Python and a Flask backend with a React frontend. 
To demonstrate \name{}'s interface and interaction, we present a user scenario featuring Alice (\autoref{fig:UserScenario}), a visual designer who used \name{} to create a Chinese-style illustration around ``environmental protection''.


\subsubsection{Get Suggested Cultural Symbols and Search Chinese Paintings}
Alice begins by brainstorming related symbols. \raisebox{-.18\height}{\includegraphics[width=0.855cm]{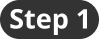}} She inputs the design theme ``environmental protection'' into the Symbol Association Panel 
and receives suggested cultural symbols, including mountains, water, auspicious clouds, birds, and deer. She then explores the cultural contexts and poetic examples related to these keywords. 
While reading the verse ``In the thick woods a deer is seen at times; Along the stream I hear no noonday chimes'', she identifies the deer as an intriguing symbol that represents harmony with nature. 
\raisebox{-.18\height}{\includegraphics[width=0.86cm]{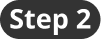}} Inspired by this, she searches for Chinese paintings featuring deer and drags several images onto the moodboard.
She then engages with these images with tags for further insight. 
\penguin{
To avoid overwhelming users, \name{} displays the tags about \textcolor{symbol}{cultural symbols}, \textcolor{emotion}{emotions}, \textcolor{composition}{compositions}, and \textcolor{style}{styles} (\autoref{tab:designspace}) that designers often learn from Chinese paintings (\autoref{sec:four_dimensions}). 
Tags about brushstroke and color tone only visible if users click the ``+'' button in the end of the displayed tags. 
To retain the tags' semantic meanings, we did not display the higher-level, more abstract categories (\eg Soul of Human Life) of the tags about cultural symbol and emotion. 
Though \name{} does not perform well in recognizing composition, and emotion is subjective, the related tags are displayed as they could still provide words that inspire and express users' thoughts. 
}
\raisebox{-.18\height}{\includegraphics[width=0.86cm]{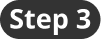}}
Among the images, one of them is tagged with ``Nine-Colored Deer''.
Fascinated by this new concept, Alice sends the concept to the Image Library 
for further exploration.
She activates the online search mode to access a broader range of resources of the Nine-Colored Deer.

\subsubsection{Analyze the Design Space of Chinese Paintings and Form Ideas}
\raisebox{-.18\height}{\includegraphics[width=0.86cm]{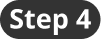}}
Alice continues her exploration, initially focusing on blue-green landscapes. Over time, her focus shifts to other types of landscape cultural symbols.
From the tags associated with one particular landscape painting, she notices the keyword ``Loneliness'', which stands out as distinct from the earlier blue-green landscapes. 
\raisebox{-.18\height}{\includegraphics[width=0.86cm]{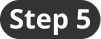}}
Intrigued, she conducts further searches with the concept of ``Loneliness''.
The retrieved images display a dramatic shift in style, with tags like ``Dry Landscape''.
In these new images, Alice finds features such as ``bare trees'' and ``gnarled rocks'' particularly interesting, 
motivating her to explore several related images. 
\subsubsection{Generate Visual Representations of the Ideas}
Alice reviews the explored content on the moodboard but found it challenging to imagine specific ideas and their corresponding visuals. 
\raisebox{-.18\height}{\includegraphics[width=0.865cm]{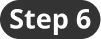}}
So, she selects several tags that interested her, such as ``Nine-Colored Deer’’, ``Blue-Green Landscape'', and ``Dry Landscape''. 
Before generating, she reviews the tags and realizes that she is uncertain whether the ``deep perspective'' is appropriate. 
So, she clicks on the tag and finds the Chinese paintings with this tag highlighted in the moodboard (images in \raisebox{-.18\height}{\includegraphics[width=0.86cm]{sections/fig/icon_step/step4.png}}).
She clicks the ``Generate'' button to quickly generate sketches to test their potential combinations. 
\raisebox{-.18\height}{\includegraphics[width=0.86cm]{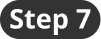}}
One of the sketches stood out: it depicts a dramatic contrast between a barren, rocky mountain in the distance and a vibrant village with peach blossoms in the foreground. 
Alice finds this contrast intriguing, as it unexpectedly combines elements of blue-green landscapes with dry landscapes.
She drags this sketch onto the moodboard to save it and manually adds the tag ``Yin-Yang Contrast'' to record her inspiration.
\raisebox{-.18\height}{\includegraphics[width=0.86cm]{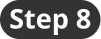}}
Then, with a clearer idea in mind,  she enters her refined concept into the input box, \raisebox{-.18\height}{\includegraphics[width=0.86cm]{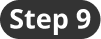}} adds one image to the collection,
and uses it as a reference for optimizing the visual style. \raisebox{-.18\height}{\includegraphics[width=1cm]{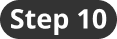}} The generated image meets her expectation, and she decides to proceed with her design based on this foundation.
\pzh{
\subsection{\penguin{Implementation and Qualitative Validation}} \label{sec:qualitative_validation}
}

\penguin{
In this subsection, we provide details on the implementation and evaluation of AI-generated content used in \name{}.
We invited three raters with training in Chinese paintings (\ie S04, S06 in \autoref{tab:formativesusertable} and S11 in \autoref{sec: dataset}) to evaluate the quality of the AI generated content, including cultural symbols, images, design intentions, and poems. 
All items were rated on a standard 5-point Likert scale, with 1 for strongly disagree and 5 for strongly agree. 
The mean scores of three raters are reported in 
\autoref{tab:output_score}. 
\shiwei{All the prompts \wu{and generated outputs used for evaluation} are provided in the supplementary materials.}
\wu{Detailed evaluation procedure and inter-rater agreements are reported in \autoref{sec: evaluationappdix}.}}
\\
\subsubsection{\wu{Cultural Symbol Association}}
We prompt GPT-4o to recommend symbols based on the design theme.
Additionally, we also prompt AI to generate literary references related to the symbol, drawing from classical Chinese cultural texts (\eg poetry) and explaining their relevance to the cultural context. 
\penguin{
Raters were asked to evaluate the \textit{relevance} and \textit{diversity} of the \wu{20 sets of} generated cultural symbols in relation to their corresponding design themes. 
As shown in \autoref{tab:output_score}, they perceived the cultural symbols as being of high quality. 
}
\subsubsection{\wu{Image Search}}
\name{} leverages Elasticsearch
for indexing and searching a large number of data repositories. 
Images are stored on a server and indexed in Elasticsearch using their URLs and AI-annotated tags.
When a query is made, the system retrieves relevant images by matching these keywords and detailed description.

\subsubsection{\penguin{Image Generation}}
For image generation, we utilized the user-selected tags to prompt GPT-4o to craft a comprehensive design intention, integrating domain knowledge 
to refine the descriptions of Chinese painting style terms with more precise and specific visual characteristics.
Subsequently, MidJourney \penguin{API} was employed to generate the visual outputs based on the crafted design intention.
Then, the generated images were provided to GPT-4o mini, which was prompted as an expert in matching Chinese poems to Chinese paintings. 
It was instructed to compose a poem based on the image's visual elements (\ie cultural symbols, composition, color, brushwork, and emotion) and to return an existing poem if appropriate, or generate a new one otherwise.

\begin{figure*}[htbp]
    \centering
\includegraphics[width=1.9\columnwidth]{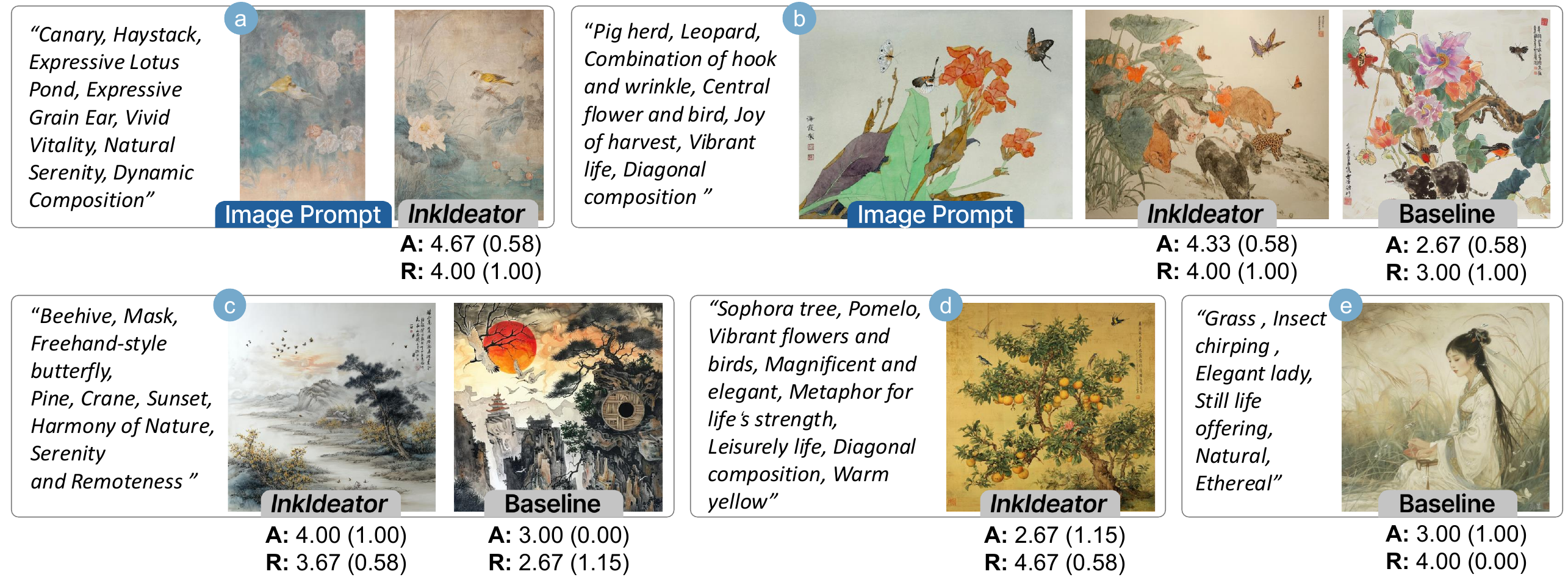}
    \caption{
    \wu{
        \textbf{Success and failure of AI-generated Images:} (a), (b)-\textit{InkIdeator} and (c)-\textit{InkIdeator} are perceived to have high alignment with the aesthetics of Chinese painting. (b)-Baseline, (c)-Baseline, (d) and (e) are perceived to have low alignment with the aesthetics of Chinese painting. Each letter at the bottom of images stands for: A: \textit{Aesthetics}, R: \textit{Relevance}. Each number indicates the mean score (SD).
    }
    }
    \label{fig:imageevaluation}
\end{figure*}

\penguin{
Raters were asked to evaluate generated images, design intentions, poems for 20 sets of randomly chosen tags, with five sets additionally having image prompts selected by the authors to simulate real use cases of \name{}. For each set, we asked raters to evaluate the \textit{relevance} between design intention and the tags, the \textit{relevance} between the first generated image and the tags, their \textit{preference} on the first generated image and its alignment of the \textit{aesthetics} of Chinese painting, the \textit{diversity} of all \wu{three} generated images, 
and the \textit{relevance} between the generated poem and the first image. 
These items are adapted from \cite{hou2024c2ideas}. 
We also had a baseline condition for generating images by directly prompting Midjourney using tags and a keyword ``Chinese painting''. 
}
\penguin{
\wu{As shown in \autoref{tab:output_score},} participants perceived the design intention 
and the generated image \wu{in \name{} condition} 
quite relevant to the tags. 
\wu{The generated images in \name{} condition are perceived more relevant to the tags compared to the baseline condition 
($Z = -2.108, p = 0.035$).
This can account for our crafted prompt transforms the tag combination into a more comprehensive design intention. S06 noted on \autoref{fig:imageevaluation}c-\name{}, ``\textit{Although I feel that the combination of tags is not very harmonious, the image is harmonious}''. She also noted on \autoref{fig:imageevaluation}c-Baseline, ``\textit{overall composition is a little weird}''.}
\autoref{fig:imageevaluation}b-Baseline features an animal with a bear-like body and leopard spots on its tail.
The generated 
images \wu{for each set of tags} have moderate diversity \wu{in both conditions}.
The generated poem was considered relevant to its image. 
Raters also showed moderate preference on the generated images 
and perceived them as reasonably aligned with the aesthetics of Chinese painting 
\wu{in both conditions}. 
}

\begin{table}[]
\caption{
    \wu{\textbf{User ratings of the AI output.}Participants' evaluation (N = 3) on the generated cultural symbols, design intention, images, and poems. 
    All items are measured using a standard 5-point Likert scale (1 - strongly disagree; 5-strongly agree).
    Items about generated images were analyzed using Wilcoxon signed rank test. Note: $*:p<0.050$.
    } 
    }
    \label{tab:output_score}
\scalebox{0.85}{
\begin{tabular}{@{}llllcc@{}}
\toprule
\multicolumn{1}{c}{\multirow{2}{*}{Output}} & \multicolumn{1}{c}{\multirow{2}{*}{Item}} 
& \multicolumn{1}{c}{InkIdeator} & \multicolumn{1}{c}{Baseline} & \multicolumn{2}{c}{Statistic} \\ \cmidrule(l){5-6} 
\multicolumn{1}{c}{}   & \multicolumn{1}{c}{}  & \multicolumn{1}{c}{Mean (SD)}  & \multicolumn{1}{c}{Mean(SD)}  
& Z  & $p$                         \\ \midrule
\multirow{2}{*}{Cultural Symbol}  & Relevance    & 4.57 (0.59)  & \multicolumn{1}{c}{-}  & -   & -  \\
                                & Diversity   & 4.47 (0.85)  & \multicolumn{1}{c}{-}       & -     & -   \\ \midrule
Design Intention                & Relevance   & 4.73 (0.61) & \multicolumn{1}{c}{-}        & -     & -   \\ \midrule
\multirow{4}{*}{Image}          & Relevance   & \textbf{4.10 (0.82)}  & 3.73 (0.92)  & \multicolumn{1}{r}{-2.108} & \multicolumn{1}{l}{{0.035}\textsuperscript{*}} \\
                                & Preference  & 3.45 (1.19)  & 3.22 (1.14)  & \multicolumn{1}{r}{-1.437} & \multicolumn{1}{l}{0.151} \\
                                 & Aesthetics & 3.40 (1.22)  & 3.42 (1.05)  & \multicolumn{1}{r}{0.093}  & \multicolumn{1}{l}{0.926} \\
                                & Diversity   & 3.68 (1.26)  & 3.63 (0.86)  & \multicolumn{1}{r}{-0.130} & \multicolumn{1}{l}{0.896} \\ \midrule
Poem                            & Relevance   & 4.05 (1.06)  & \multicolumn{1}{c}{-} & -                          & -  \\ \bottomrule
\end{tabular}
}
\end{table}
\penguin{
\autoref{fig:imageevaluation} shows some qualitative examples of success and failure. 
The raters' feedback highlight two insights. 
First, the relevance among the tags would affect the downstream generated Chinese paintings, though raters could find their own ways to interpret the relevance. 
S04 noted, ``\textit{Putting leopards and pigs together works too; it's pretty surreal and it can represent the harmony of life, since predators and prey can coexist peacefully.}''(\autoref{fig:imageevaluation}b). 
S11 noted, ``\textit{I find this set of words particularly well-combined. Words like `pine, crane, sunset, harmony, and tranquility' evoke an atmosphere of an ideal pursuit. At the same time, the butterfly and the mask can symbolize the spirit of freedom and the confinement of the soul, respectively. Through this contrast, the longing for such an atmosphere is further highlighted.}''(\autoref{fig:imageevaluation}c). 
}

\penguin{
Second, \wu{
we found that including a Chinese painting as an image prompt can help generating images aligned with the aesthetics of Chinese paintings (\autoref{fig:imageevaluation}a, b), but}
the generated images can sometimes contain mistakes. 
S04 commented on \autoref{fig:imageevaluation}b-\name{} , ``\textit{missing symbols (birds)}''.
S11 noted that the colors in \autoref{fig:imageevaluation}d are ``\textit{too heavy and not light enough, unlike those in traditional Chinese paintings, and instead resemble a Western color scheme}''.
S04 noted on \autoref{fig:imageevaluation}e, ``\textit{The face is too prominent, more like an illustration rather than a traditional Chinese painting}''. 
}
\section{User Study}
\begin{table*}[]
\caption{\wu{Demographic of participants in the user study.}}
\label{tab: participant}
\scalebox{0.73}{
\begin{tabular}{@{}clcllcl@{}}
\toprule
ID                         & \multicolumn{1}{c}{Gender} & Age & \multicolumn{1}{l}{Major}                 & \multicolumn{1}{c}{Chinese-style Design Experience} & Study Design Duration & \multicolumn{1}{c}{Chinese Painting Expertise (Perceived Score)}  \\ \midrule
{P01} & Female  & 24  & Industrial Design  & \begin{tabular}[c]{@{}l@{}}Have designed Mid-Autumn Festival \\ traditional Chinese style posters, etc.\end{tabular}                     & 5 years               & \begin{tabular}[c]{@{}l@{}}No formal professional course training, \\ but have some related knowledge.(3)\end{tabular}                             \\
{P02} & Female                     & 19  & Industrial Design                                    & Have designed traditional Chinese style poster.                                                                                          & 2 years               & \begin{tabular}[c]{@{}l@{}}No formal professional course training, \\ but have some related knowledge.(2)\end{tabular}                             \\
P03                        & Female                     & 24  & Visual Communication Design                          & Have designed Ink style posters.                                                                                                         & 7 years               & \begin{tabular}[c]{@{}l@{}}Have some knowledge about traditional \\ Chinese painting.(2)\end{tabular}                                              \\
P04                        & Female                     & 20  & Information Art Design                               & \begin{tabular}[c]{@{}l@{}}Have designed original IP character \\ design for the "Qingluan Yin" series.\end{tabular}                     & 3 years               & {\color[HTML]{212121} \begin{tabular}[c]{@{}l@{}}Have knowledge about composition \\ in traditional Chinese painting.(3)\end{tabular}}             \\
P05                        & Female                     & 21  & {\color[HTML]{212121} Digital Media Art}             & \begin{tabular}[c]{@{}l@{}}Have designed traditional Chinese style tissue \\ box, cartoon avatars, etc.\end{tabular}                     & 3 years               & {\color[HTML]{212121} \begin{tabular}[c]{@{}l@{}}Have taken traditional Chinese painting \\ classes, can paint but not proficient(3)\end{tabular}} \\
P06                        & Female                     & 23  & {\color[HTML]{212121} Dyeing and Weaving Art Design} & \begin{tabular}[c]{@{}l@{}}Have designed many traditional Chinese style \\ scarves, bedding, etc.\end{tabular}                           & 5 years               & Have never been exposed to it.(2)                                                                                                                  \\
P07                        & Female                     & 23  & {\color[HTML]{212121} Digital Media Art}             & \begin{tabular}[c]{@{}l@{}}Have completed projects related to ancient \\ architecture (3D model building, related posters).\end{tabular} & 4 years               & \begin{tabular}[c]{@{}l@{}}Exposed to traditional Chinese painting \\ in elementary school.(2)\end{tabular}                                        \\
P08                        & Female                     & 20  & Painting                                             & Have designed illustrations with the Dunhuang theme.                                                                                     & 4 years               & \begin{tabular}[c]{@{}l@{}}Have done traditional Chinese painting \\ in school courses.(4)\end{tabular}                                            \\
P09                        & Female                     & 18  & New Media Art and Design                             & \begin{tabular}[c]{@{}l@{}}Have designed traditional Chinese style \\ collage illustrations, posters, etc.\end{tabular}                  & 1 year                & \begin{tabular}[c]{@{}l@{}}Studied traditional Chinese painting for\\ three years.(3)\end{tabular}                                                 \\
P10                        & Female                     & 24  & Chinese Painting                                     & \begin{tabular}[c]{@{}l@{}}Have designed posters for traditional Chinese \\ painting exhibitions.\end{tabular}                           & 1 year                & \begin{tabular}[c]{@{}l@{}}Studied traditional Chinese painting for \\ five years.(4)\end{tabular}                                                 \\
P11                        & Female                     & 23  & Design Studies                                       & \begin{tabular}[c]{@{}l@{}}Have designed Chinese-style posters, picture books,\\  and illustrations.\end{tabular}                        & 5 years               & \begin{tabular}[c]{@{}l@{}}No formal professional course training, \\ but have some related knowledge.(3)\end{tabular}                             \\
P12                        & Male                       & 23  & Visual Communication Design                          & \begin{tabular}[c]{@{}l@{}}Have designed public welfare and traditional Chinese \\ style posters.\end{tabular}                           & 2 years               & Level 6 in traditional Chinese painting.(3)                                                                                                        \\ \bottomrule
\end{tabular}
}
\end{table*}
\begin{figure*}[]
    \centering
    \includegraphics[width=1.85\columnwidth]{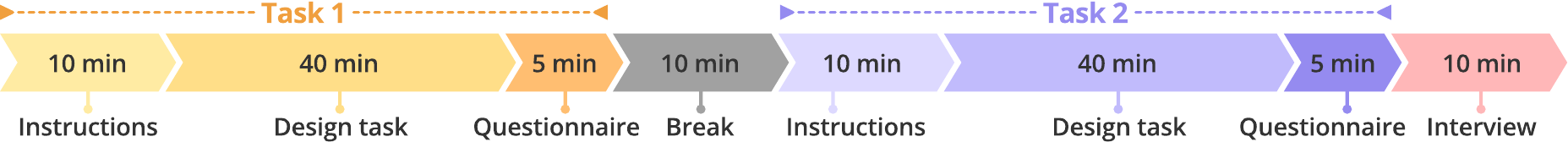}
    \caption{\wu{\textbf{User study procedure.} The tasks and interfaces (\ie Baseline and \textit{InkIdeator}) were counterbalanced using Latin Square, and a post-interview was conducted. The study lasted approximately 2 hours.}}
    \label{fig:user study}
\end{figure*}
\pzh{
We conducted a within-subjects user study with 12 participants to evaluate the effectiveness, user experience, and usefulness of \name{} for supporting Chinese-style visual design ideation with Chinese paintings. 
}
\subsection{Baseline} 
\pzh{The most unique features of \name{} are centered around the extracted keywords of cultural symbols, emotions, compositions, and styles in Chinese paintings, which are displayed and interactive in the MoodBoard (\autoref{fig: interface}C) and Image Generation Panel (\autoref{fig: interface}D).
Besides, \name{} is unique regarding our developed GenAI pipeline to support the Chinese-style visual design ideation process.
To examine the values of these unique features, we implemented a baseline system (\autoref{fig:baseline} in Appendix) that provides an Image Search Panel using the same annotated dataset of Chinese paintings and a MoodBoard similar to \name{}, except that the annotated keywords are not displayed. 
Participants can also use GenAIs (\ie ChatGPT, Midjourney) but in their original platforms. 
}
\subsection{Participants}
We recruited 12 participants (denoted as P01-P12, 11 females, 1 male; age $M = 21.83$ and $SD = 2.03$) through an online recruitment posting in a university, with inclusion criteria of holding a degree in design or art and having prior experience in Chinese-style visual design (\eg Ink painting-style poster, Chinese-style illustration).
Ten were students majoring in design — five at the graduate level and five at the undergraduate level. 
The remaining two participants were students majoring in art, with one graduate and one undergraduate student.
\penguin{
On a 5-point Likert scale (1: no knowledge, 5: extensive knowledge), four participants self-reported limited knowledge (2), six had some knowledge (3), and two had good knowledge (4) and reported experiences in painting or appreciating Chinese paintings. 
}
\penguin{
All participants were not specialized in Chinese paintings, and their background are included in \autoref{tab: participant}.
}
\subsection{Task and Procedure}



\autoref{fig:user study} illustrates the procedure of our user study conducted remotely via a video-conferencing application. 
Participants were asked to complete design ideation tasks twice in two settings, \ie \name{} and baseline. 
\wu{We counterbalanced the tasks and interfaces using Latin Square.}
The task was to create a Chinese-style illustration for the cover of a book with the theme ``Environmental Protection'' or ``Aerospace''. 
These two themes are two hot topics in modern society, making it a non-trivial Chinese-style design task as the participants need to consider both cultural and modern elements.
Participants in two conditions were not permitted to use external search engines or tools.
Before starting each task, we introduced the basic functions of the assigned system and allowed participants 3–5 minutes to familiarize themselves with the interface.
Participants were then given approximately 35 minutes (no more than 40 minutes) to complete the ideation task using the assigned tool.
After completing each task, participants were asked to verbally articulate their ideation outcomes, using visual references (\eg Chinese paintings in the moodboard, generated images) to support idea presentation.
They were encouraged to explain the cultural elements in their ideas and their ideation process.
Then, participants completed a questionnaire about their experience during the ideation process.
Participants were given a 10-minute break between each task.
After finishing both two tasks, a semi-structured interview was conducted, focusing on their perceptions of the used tools, feedback on the specific features, and suggestions for improving \name{}.
The study was conducted for about 2 hours, and we compensated participants with 120 CNY (about 16.8 USD).
\subsection{Measures}





\pzh{We instructed participants that the ideation outcome is not a generated image nor the completed visual design using their design tools. Instead, the outcome is an idea that they should verbally communicate to the experimenter, using explored Chinese paintings or generated examples as references.}
Using references to communicate the design ideas with clients is a common practice, especially under time constraints \cite{wang2025aideation}.
For ideation outcome, we transcribed participants' verbal responses and compiled a comprehensive document that presents their ideas with references and descriptions.
We recruited two external raters with graduate degrees in design and had 3 and 2.5 years of experience teaching design each.
Both raters had Chinese-style design experience. 
For example, one had participated in the art design of
``Nobody'' \footnote{\url{https://en.wikipedia.org/wiki/Nobody_(2025_film)}}.
We asked them to evaluate the clarity and appeal of the idea description on a 7-point Likert scale (Q1, 1 - lowest clarity and appeal, 7 - highest clarity and appeal).
A total of 24 ideation outcome (2 conditions $\times$ 12 participants) were evaluated.
If the score difference between the two experts exceeded two points, they were required to discuss and decide whether to adjust their scores.
The final score for each ideation outcome was the average of their scores after the discussion.

As for user experience, we utilized the Creativity Support Index CSI \cite{cherry2014quantifying} (Q2-Q7) to measure the degree of creativity support (1 - strongly disagree, 7 - strongly agree) and NASA-TLX \cite{hart2006nasa} (Q11) to evaluate perceived workload (1 - least workload, 7 - highest workload). 
We also included three 7-point Likert scale questions focused on the three design goals (Q8-10). 
In the \name{} condition, we also asked participants to rate the usefulness of each panel (Q12-15).
These items were rated on a standard 7-point Likert scale (1 - strongly disagree, 7 - strongly agree).
\subsection{Data Analysis}
Considering the small sample size (N=12) and the ordinal nature of the data, we employed a Wilcoxon signed-rank test \cite{woolson2005wilcoxon} to compare the self-reported items between \name{} and the baseline tool.
For the idea description quality, we confirmed the data normality using Shapiro-Wilk test \cite{shapiro1965analysis} and
conducted a paired sample t-test \cite{ross2017paired} to compare the differences of idea description quality between the \name{} and baseline conditions.
For the interview recordings, two of the authors transcribed them into text and conducted a thematic analysis \cite{clarke2017thematic}.
They first familiarized themselves by reviewing all the text scripts independently.
After several rounds of coding with comparison and discussion, they finalized the codes of all the interview data.

\begin{figure*}[]
    \centering
\includegraphics[width=0.85\textwidth]{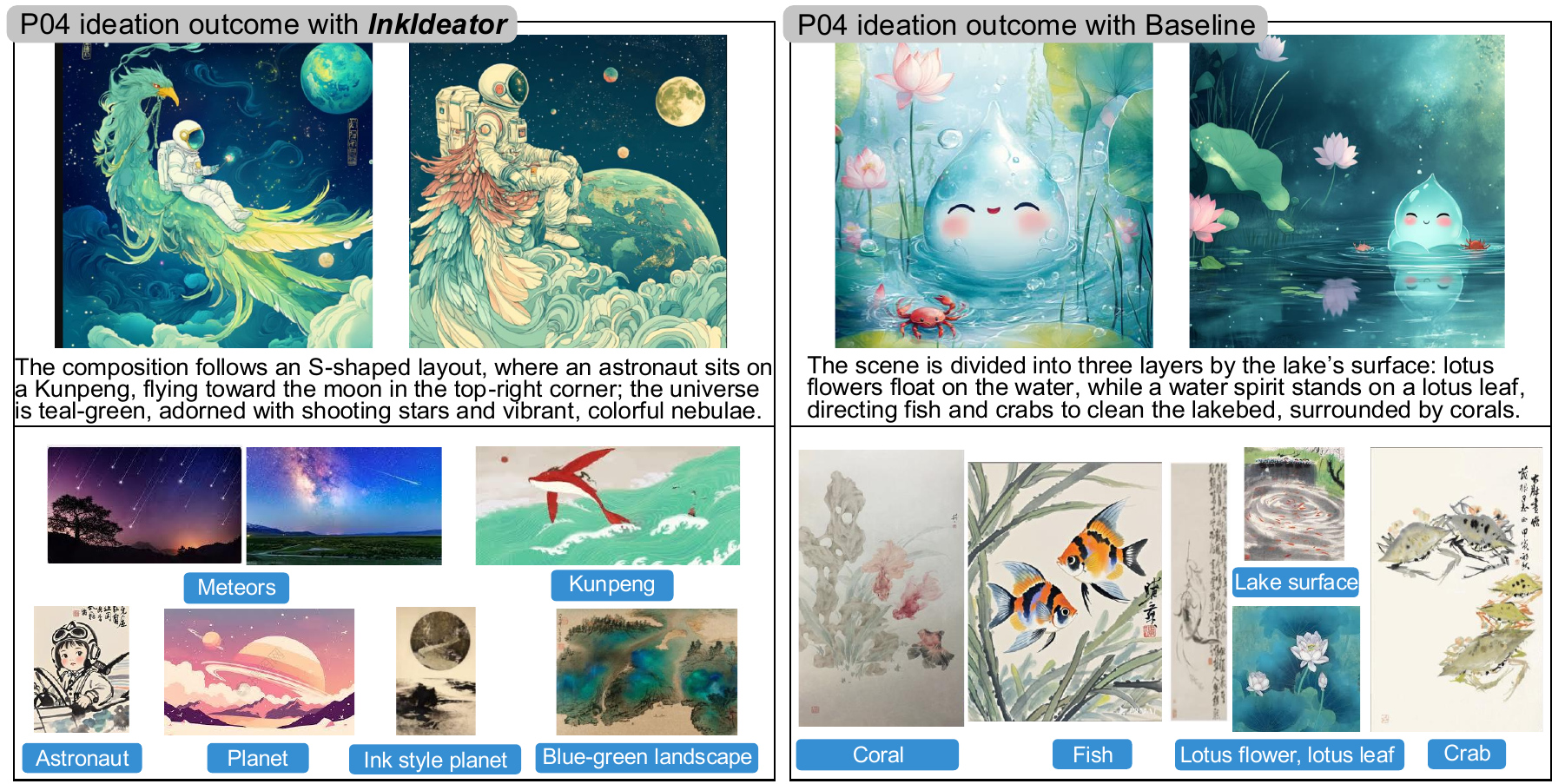}
    \caption{P04 ideation outcomes and sources of inspiration during the ideation process.}
    \label{fig:P04}
\end{figure*}
\begin{figure*}[]
    \centering
\includegraphics[width=0.85\textwidth]{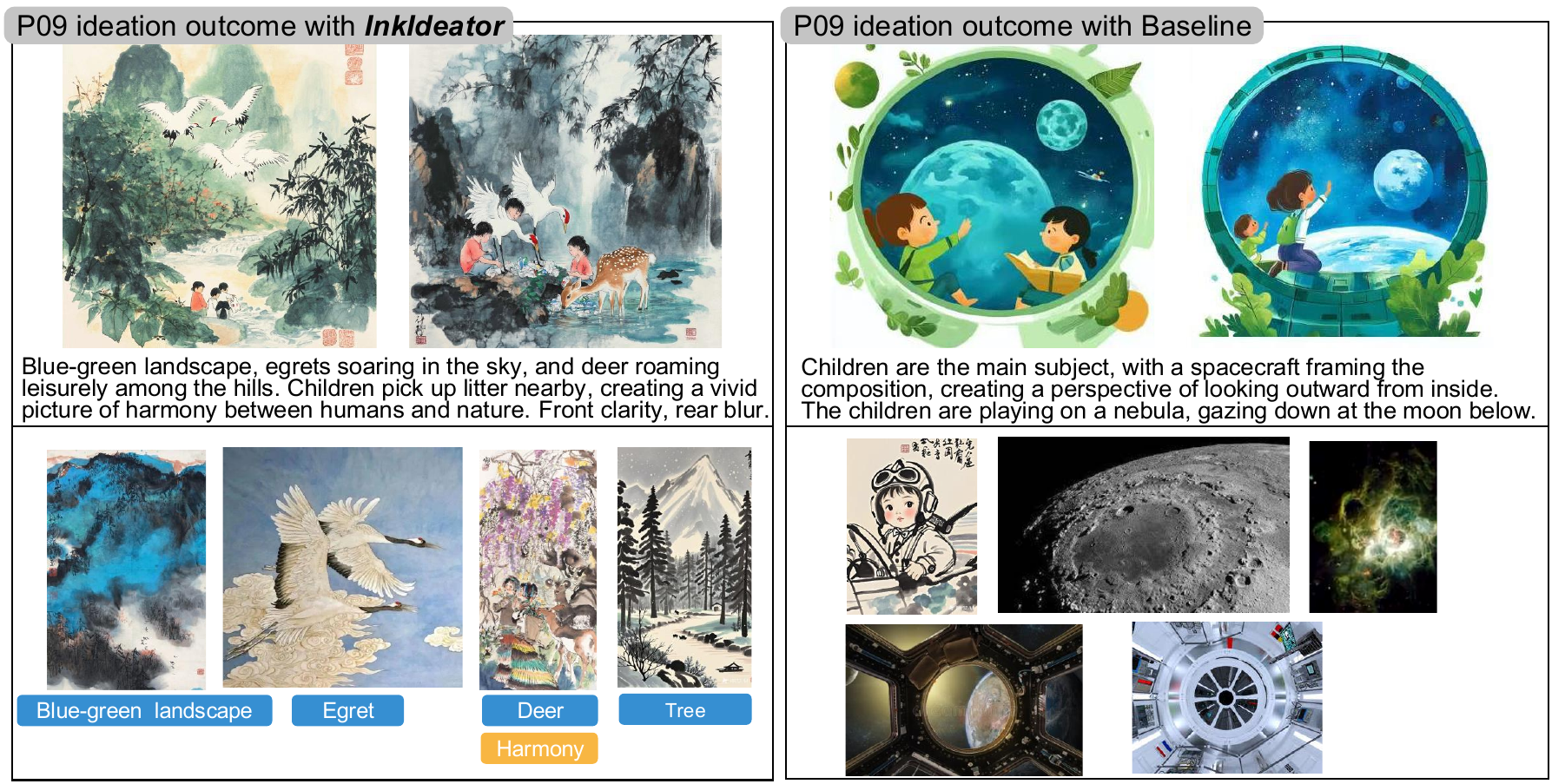}
    \caption{P09 ideation outcomes and sources of inspiration during the ideation process.}
    \label{fig:P09}
\end{figure*}
\begin{figure*}[]
    \centering
\includegraphics[width=0.9\textwidth]{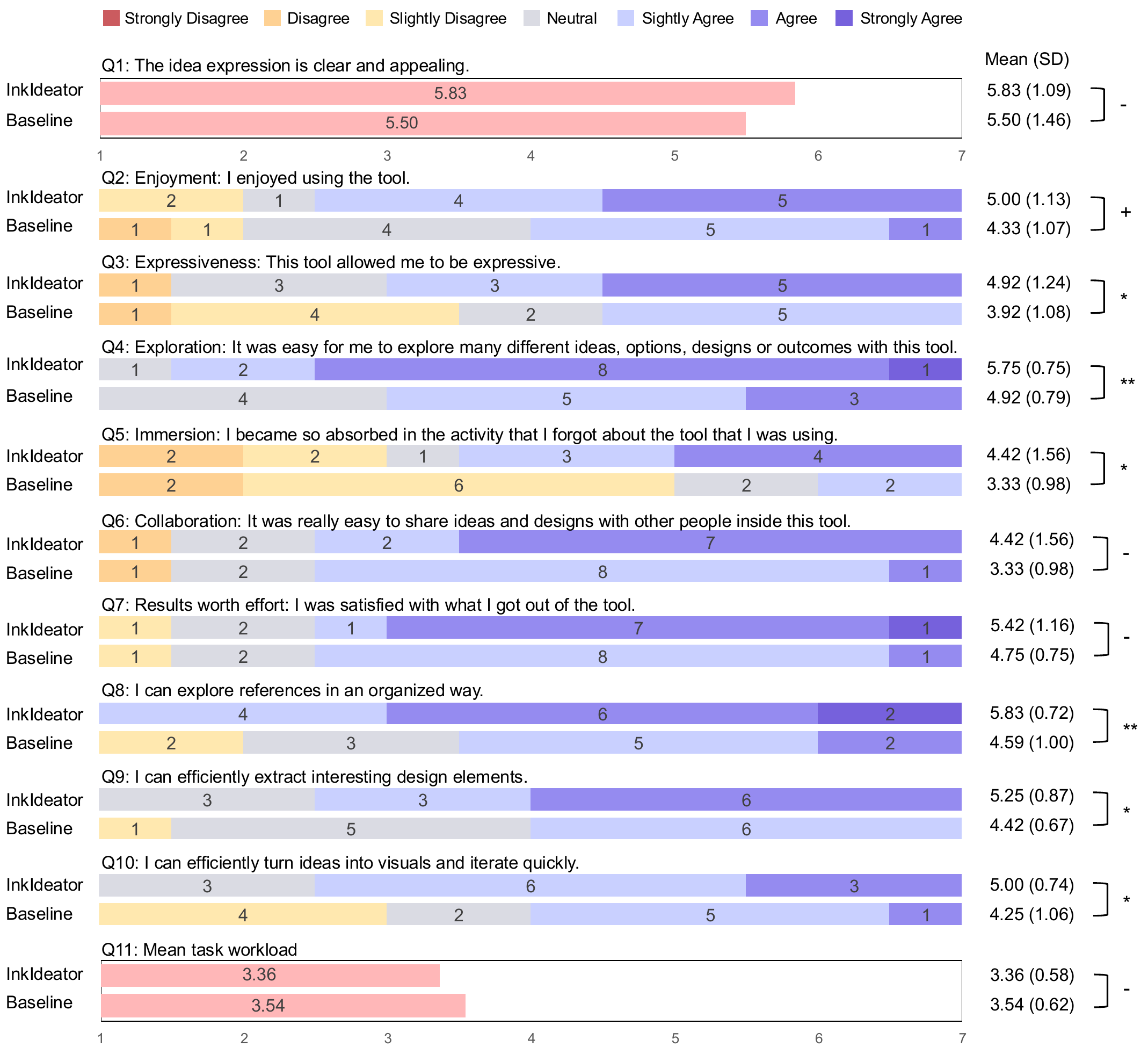}
    \caption{The experiment's statistical results about baseline and \name{} system. 
    Idea description was analyzed using paired sample t-test, while others using Wilcoxon signed rank test.
     Note: $***:p <0.001; **:p<0.01; *:p<0.05; +:0.05<p<0.1; -:p>0.1$ ; within-subjects; N = 12.}
    \label{fig:result}
\end{figure*}
\begin{figure*}[]
    \centering
\includegraphics[width=0.78\textwidth]{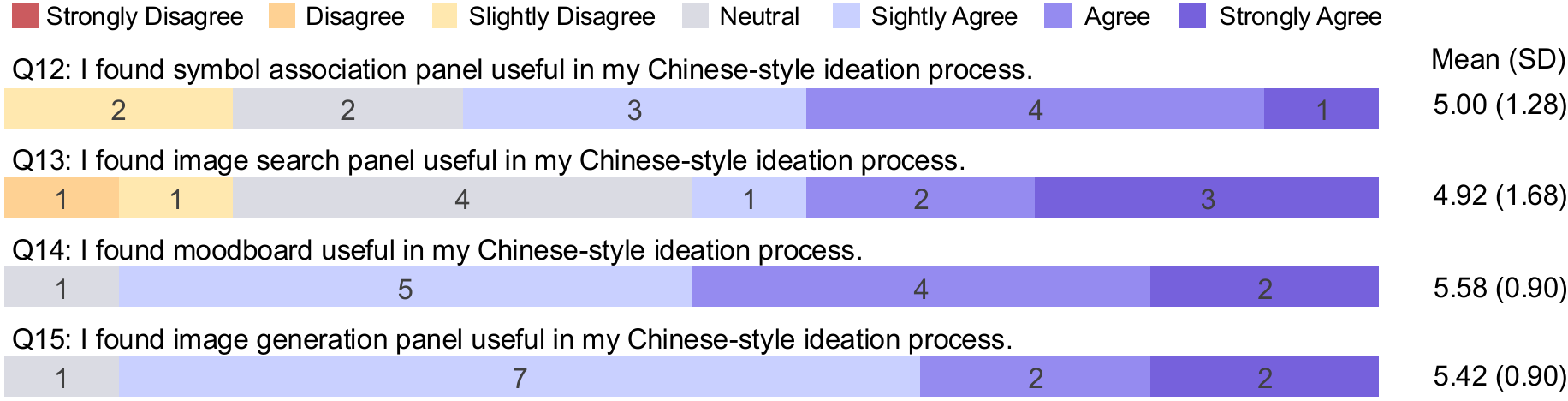}
    \caption{The participants score about usefulness of \textit{InkIdeator} interface. }
    \label{fig:function}
\end{figure*}
\section{Results} \label{sec: results}


This section first presents the quantitative results (\autoref{fig:result}), followed by the qualitative findings on whether and how \name{} facilitated participants' ideation process. 

\subsection{Quantitative Results}

\textbf{Ideation Outcome.}
While participants' design ideas with \name{} ($M = 5.83$) generally have better clarity and appeal than the ideas with the baseline system ($M = 5.50$), the difference is not significant (Q1 in \autoref{fig:result}, $p = 0.254$). 
\pzh{The relatively high average scores in both conditions suggest that the Chinese paintings, along with GenAIs, were generally helpful for Chinese-style visual design ideations.}
\autoref{fig:P04} and \autoref{fig:P09} showcase examples of two participants' (P04, P09) ideas in two conditions, which include the textual description as well as the referenced generated images and explored examples. 
\pzh{
Specifically, though the displayed Chinese paintings in the baseline moodboard do not contain generated tags, P4 actively tagged her ideas on the explored paintings (\autoref{fig:P04} bottom right).
These examples of ideation outcome indicate that participants actively learned from Chinese paintings to form their visual design ideas in our study.
}

\textbf{Ideation Experience.}
\pzh{While the outcome ideas look comparably clear and appealing, participants perceived that they received better creative support from the \name{} than the baseline system (Q2-Q7 in \autoref{fig:result}).
Specifically, participants felt that \name{} was more enjoyable to use (Q2: $Z = -1.809, p = 0.070$), 
allowed them to be significantly more expressive (Q3: $Z = -2.360, p = 0.018$), 
made it significantly easier to explore many different ideas, options, designs, or outcomes (Q4: $Z = -2.887, p = 0.004$), 
led to significantly more immersive experience (Q5: $Z = -2.228, p = 0.026$),
made it more easier to share ideas and designs with other people \peng{(Q6: $Z = -1.430, p = 0.153$)}, 
and made them generally more satisfied (Q7: $Z = -1.331, p =0.183$). 
Participants with \name{} also perceived slightly less mean task workload in the ideation task (Q11: $Z = 0.449, p =0.653$), though the difference is not significant. 
Besides, \name{} significantly outperformed the baseline system in helping participants explore references in an organized way (Q8: $Z = -2.724, p =0.006$), efficiently extract interesting design elements (Q9: $Z = -2.153, p =0.031$), and efficiently turn ideas into visuals and iterate quickly (Q10: $Z = -2.310, p =0.021$).
These findings indicate that \name{} successfully reaches its design goals (\autoref{sec:design_goals}).
}

\textbf{Usefulness of Functions.}
As shown in \autoref{fig:function}, overall, participants found the Symbol Association ($M = 5.00, SD = 1.28$), Image Library ($M = 4.92, SD = 1.68$), Moodboard ($M = 5.58, SD = 0.90$), and Image Generation ($M = 5.42, SD = 0.90$) panels useful in the ideation process.
\subsection{Qualitative Results}


\subsubsection{InkIdeator supports effective searching and analyzing Chinese paintings as well as visualizing the ideas that integrate multiple dimensional keywords.}\label{sec:qualitative_finding_1}
Participants reported that the recommended cultural symbols and AI-annotated design space effectively assisted them in exploring Chinese paintings.
``\textit{At the beginning, it recommended some concrete concepts, with which I can further search for related images and visualize the scene.}'' (P11, F, 23).
P10 (F, 24) also mentioned, ``\textit{I usually browse through a large number of images to get inspiration. I think this search function with the label is very useful, as it allows me to quickly find what I want.}''
We also found that adding annotations \pzh{to the paintings in the moodboard}
can help participants organize their thoughts, and come up with new ideas.
P01 (F, 24) noted, ``\textit{These labels are very convenient and allow me to ideate in an organized manner, quickly categorizing the images. In the baseline system, the images were presented without these labels, making the process feel disorganized and chaotic, and I would need to classify them myself first.}''
P12 (M, 23) added, ``\textit{When we design, we usually need to brainstorm possible concepts, and the display of these tags can help me associate related concepts.}''
They also mentioned that choosing different dimensional keywords attached to the explored examples made it easy to generate images.
``\textit{I feel that it (InkIdeator) understands the styles of Chinese painting well. When I mentioned techniques like 'baimiao' (fine line drawing) or blue-green landscape, it can generate images that reflect these characteristics, making it more convenient than tweaking prompts myself.}'' (P09, F, 18).
``\textit{This method of selecting labels helps me organize my thoughts and streamline the generation process}'' (P12, M, 23).

\subsubsection{InkIdeator supports users in integrating cultural symbols and emotions of Chinese paintings into their ideas.}\label{sec:qualitative_finding_2}
Participants mentioned that \name{} assisted them in discovering ideas from cultural perspectives and reflecting on the emotions conveyed in their visual design ideas.
``\textit{The recommended symbol suggested Chang'e flying to the moon, which inspires me to think about the idea of a rocket flying to the moon as a whole.} (P04, F, 20)
``\textit{I saw an image of a tree that felt vibrant and full of life. It made me think about how the tree could connect to human life itself and express vitality.}'' (P12, M, 23)
``\textit{I think the poem next to the generated image serves as a kind of summary of the emotion or spirit, helping me feel the scene and reflect on my idea.}'' (P11, F, 23)

\subsubsection{InkIdeator supports users in exploring diverse ideas.} \label{sec:qualitative_finding_3}
We found that \name{} can assist designers in comprehensively exploring the design space, inspiring their own creativity rather than relying heavily on GenAIs.
When using the baseline system, participants tended to focus on a single idea, continuously modifying the AI-generated results rather than engaging in broader exploration.
``\textit{At first, I thought of an ancient poem, which roughly means: `I dare not speak too loudly, for fear of disturbing the immortals in the heavens'. I shared my initial idea with ChatGPT, asking it to help refine and enrich the details. Afterward, I used MidJourney to generate visuals based on the prompt and made iterative adjustments to better align the results with my vision.}'' \peng{(P02, F, 19)}.  
In contrast, \name{} leverages GenAIs to assist in certain steps of the ideation process, placing more emphasis on helping users to explore examples of Chinese paintings to gain inspiration.
AI-annotated dimensional values of Chinese paintings help users better focus on analyzing the design space and examining different ideas. 
``\textit{Sometimes, I may not immediately notice an image's elements or how to apply them to my work. For example, an annotation of a C-shaped composition for these cats inspired me to use a similar flying pattern for the geese in my image.}'' (P06, F, 23).
We also found that the recommended associated symbols and keyword-based generation assisted participants in exploring diverse ideas, enabling them to break free from being confined to a single concept.
``\textit{Based on my idea of the harmony between man and nature, I was inspired by some recommended cultural symbols like the Taiji, which I think could be an interesting approach to composition.}'' (P11, F, 23)
\begin{figure*}[]
    \centering
\includegraphics[width=0.92\textwidth]{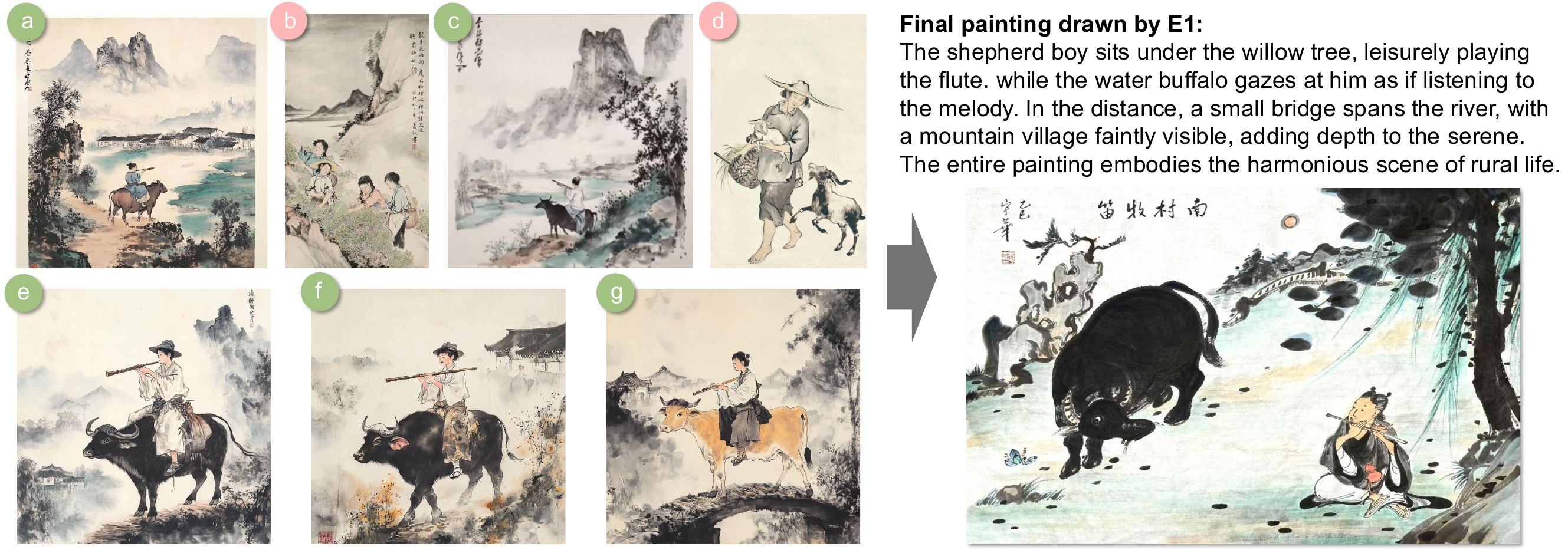}
    \caption{\wu{\textbf{Case 1: A Shepherd Boy Playing the Flute.}} E1's explored examples and the final painting drawn by him to convey his ideas (\textbf{a}, \textbf{c}, \textbf{e}, \textbf{f}, and \textbf{g} are generated images). \minor{We have obtained the artist's permission to present his painting in this paper.}}
    \label{fig:E1}
\end{figure*}
\begin{figure*}[]
    \centering
\includegraphics[width=0.92\textwidth]{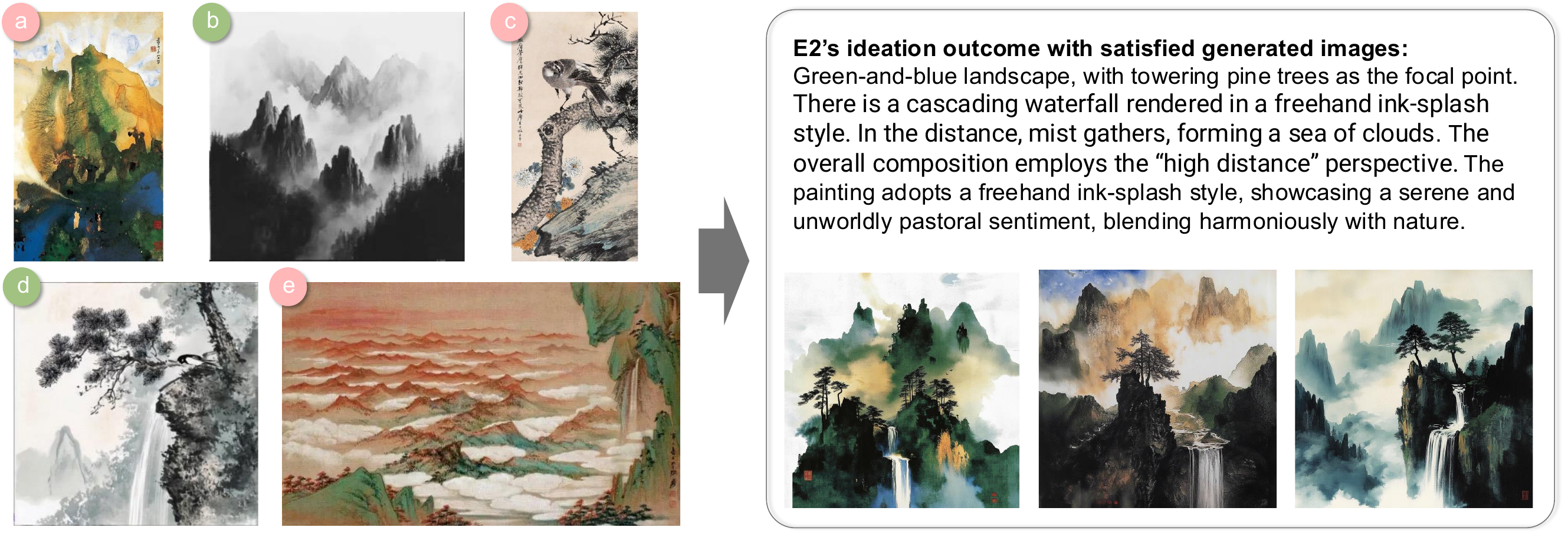}
    \caption{\wu{\textbf{Case 2: Leisurely, I See the Southern Mountains.}} E2's explored examples and generated images used to express his idea (\textbf{b} and \textbf{d} are generated images).}
    \label{fig:E2}
\end{figure*}
\rochelle{
\section{Extended Use Cases of \textit{InkIdeator} for Ideating Chinese Painting}\label{sec:expert_interview}}

\penguin{
While \name{} primarily targets ideation of Chinese-style visual design, its features of recommending cultural symbols and analyzing and generating Chinese paintings make it extensible to other creative tasks related to Chinese culture. 
To explore such potentials, we invited two experienced Chinese painters (S02, S03 in \autoref{tab:formativesusertable}, noted as E1 and E2 in this section) to use \name{} for 40 minutes to gain ideas of their ink wash Chinese paintings and think aloud in this process.}
\penguin{
\autoref{fig:E1} and \autoref{fig:E2} illustrate the painters' initial intention, explored examples, and their ideation outcomes. 
We reported their perceptions of \name{} as below. 
}


\rochelle{
\textbf{Painters seamlessly integrate cultural knowledge and visual exploration.}
Experts perceived \name{} easy to use and help them develop and refine ideas by exploring Chinese paintings with the help of AI-annotated tags and Symbol Association Panel.
E1 thought that tags can provide details for the idea.
``\textit{When I’m brainstorming, I often think about concrete elements, like imagining the shepherd holding a flute in one hand and a whip in the other to drive the cattle. These tags serve as hints for the idea, guiding me with different concepts for the painting.}''(E1).
E2 appreciated the detailed description of AI-annotated cultural symbols which provided new perspectives for expressing ideas.
For example, he wanted to draw clouds but discovered they could be depicted as a sea of clouds.
E2 also mentioned that \name{} aligned with his usual ideation process of Chinese paintings.
``\textit{For me, it is enough to have these general conceptual ideas at the beginning. We prefer to draw slowly on paper and gradually make adjustments. 
The tags in the moodboard help me gradually refine ideas across multiple dimensions.}''(E2)
}

\rochelle{
\textbf{AI-generated images serve as active inspiration triggers, and painters adopt them based on personal aesthetic judgment.}
Throughout the ideation process, whenever experts had limited ideas, they opted to generate images for inspiration.
Experts appreciated that AI-generated images, \penguin{though not perfect}, can provide a lot of inspiration.
``\textit{For a specific theme, I usually develop my own ideas, and choose techniques that suit my style. 
The AI-generated images, even if don’t perfectly align with my vision, can still offer inspiration and guidance.}'' (E1).
E2 expected the generated images to present concepts beyond his imagination.
``\textit{These generated images are more for reference and support. Most of the time, I hope they can present unexpected surprises rather than being exact replicas of my ideas.}''
They also noted some errors in the generated images.
E1 commented on \autoref{fig:E1}g, ``\textit{
I think the midground in this image feels somewhat simplistic, and the houses in the background are almost aligned along a single horizontal line. This should be avoided in Chinese painting, as it values asymmetrical balance to create visual harmony.}''
}

\penguin{Overall, these two cases showed that \name{} also aligns with the ideation process of Chinese paintings, and the design space is consistent with the concepts they consider during ideation. 
Together with the findings of our user study, \name{} has great potential to be adapted to support various creative tasks related to Chinese culture. 
}
\section{Discussion}
\pzh{
In the ideation process, designers typically experience both divergent and convergent thinking \cite{Suh24Luminate}. 
It is crucial to support designers in efficiently exploring the design space in a large amount of task-related examples \cite{dove2016argument, heape2007design, lomas2021design}.
Our formative study revealed that a lack of domain knowledge often limits designers' ability to delve deeply into the design space of Chinese paintings. 
We iteratively developed prompts to multi-modal large models for mining the cultural symbols, emotions, compositions, and styles in Chinese paintings, yielding a comprehensive design space (\autoref{tab:designspace}). 
These annotations are attached to our Chinese paintings to enable efficient example search in both \name{} and baseline system.}
Our within-subjects study shows that these tags guide designers
to analyze the design space, \wu{organize ideas with these dimensions, and} examine different ideas. 
\rochelle{User cases of ideating Chinese paintings} further demonstrate that these tags can also benefit experienced Chinese painters, helping them add details to their ideas (E1 in \autoref{sec:expert_interview}) and enhance idea expression (E2 in \autoref{sec:expert_interview}).
\subsection{Dual Roles of GenAIs on Culture Preservation}
Our studies imply that GenAIs can serve as both an amplifier and a possible distorting force in cultural continuity. 
On the positive side, 
GenAIs can lower the bar of appreciating traditional culture and creating artworks that incorporate cultural elements, fostering cultural transmission. 
For example, we had qualitatively validated that the GPT-4o can generate highly relevant cultural symbols to the design themes and relevant poems to the images (\autoref{sec:qualitative_validation}). 
Our user study further demonstrated that with GenAIs, either in \name{} or baseline condition, users did well in composing clear and appealing expressions of Chinese-style visual design ideas (Q1 in \autoref{fig:result}). 
\rochelle{User cases with Chinese painters also demonstrated AI-infused system like \name{} can facilitate artists in culture-related creative tasks.}
These findings suggest that GenAIs could also be used to support ideation of other Chinese-style design tasks, such as films (\eg Nobody) and video games (\eg Black Myth: Wukong), as designers can get inspired by the generated cultural content. 
Besides, the identified dimensions of design space guide users to think and explore with these dimensions. 
Some mis-identified tags may mislead the users, as discussed below, but may provide opportunities for users to extend thoughts and gain insights with unrelated but not random information \cite{koch2020imagesense}, which was mentioned by E1 (\autoref{sec:expert_interview}).

\penguin{
However, as reported in formative study (\autoref{sec: AI_percetion}) and qualitative validation of generated content (\autoref{sec:qualitative_validation}), GenAIs perform only moderately well in generating images that align with the aesthetics of Chinese paintings. 
Even after careful prompt engineering, GenAIs could still poorly perform in understanding certain aspects of cultural artworks, such as the compositions and brushstrokes in our case (\autoref{sec:data_annotation}). 
These could be because cultural artworks have complicated creation techniques and can be interpreted in different ways. 
For example, many techniques in Chinese painting, especially dyeing methods (\eg the eighteen dyeing techniques in \textit{gongbi} painting), focus more on the painting process, making it difficult for GenAIs to directly associate these techniques with specific visual effects. 
\wu{Another possible reason could be that the model was trained predominantly on Western images \cite{wasielewski2022beyond} which leads to bias.}
These findings align with previous work \cite{wang2025harmonycut, tao2025aifiligree}, showing GenAIs' struggle in grasping abstract cultural concepts and producing culturally aesthetic images.
Consequently, GenAIs could reinforce superficial or inaccurate representations of cultural symbols, potentially diluting the artistic and educational value of the tradition. 
This issue would be amplified if users are not familiar with cultural symbols, as they are more likely to interpret them too superficially 
to utilize these cultural concepts effectively.}

\subsection{Design Guidelines for Culture-Related Design Tools}
To encourage responsible use of GenAIs in culture-related design tools, we propose three guidelines.

\penguin{
\textbf{First, a validated knowledge base of target culture should be incorporated into the tools' design and development}. 
\wu{Our qualitative validation of the AI-generated outputs showed that manipulating domain knowledge in the prompt can improve generated images (\autoref{sec:qualitative_validation})}. 
Following our approach of mining cultural knowledge from Chinese paintings, design support tools of other cultural artworks (\eg Western oil painting) can involve domain experts to input cultural knowledge (\eg dimensions such as light and shadow treatment) and provide feedback on the AI-annotated results. 
Nevertheless, prompt engineering to GenAIs would sometimes make mistakes in analyzing and producing culture-related content, as shown in our case. 
Fine-tuning GenAIs using the manually verified data from our annotated Chinese paintings could be a potential direction for improvement.
However, the extensive and complex nature of cultural knowledge may make it unrealistic to fine-tune current models only on expanding labeled datasets \cite{wang2025harmonycut}. 
An alternative solution is to use the knowledge base to afford pedagogical elements, such as telling what specific variants of identified composition techniques, how to recognize them, and why they matter when users hover on the related tags in \name{}. 
This would train users to have an appreciation of target cultural artworks to interpret and judge the appropriateness of the generated content.
\wu{Another potential solution is to involve human in the loop, allowing users to identify and mark incorrect tags, as well as contribute their own emotions or inspirations to the images. These annotations can then be incorporated into the database for continuous improvement.}}

\penguin{\textbf{Second, a culture-related design tool powered by GenAIs should position itself as a supporter in the design process rather than a producer of directly usable outcomes}. 
\name{} is positioned as an ideation support tool that helps users explore existing Chinese paintings and visualize their potential ideas with GenAIs.
The generated images are not expected to be final design outcome but rather, together with the explored existing paintings, are references for expressing their ideas (\autoref{fig:P04} and \autoref{fig:P09}), as measured in the user study. 
In the case studies, E1 learned from the generated images, but more importantly, he input his thoughts and applied his own techniques in the final design outcome (\autoref{fig:E1}).
\minor{
\name{}'s pipeline (\autoref{fig:stages}), comprising symbols association, example-based exploration with the moodboard, and visual generation for idea reflection and representation, effectively supports the iterative nature of divergent and convergent design thinking, and could also be generalized to other types of art creation (\eg oil paintings, shadow puppetry, paper-cutting) by integrating domain-specific cultural symbols and artistic terminology, and adapting specific functions.}
Nevertheless, tools like \name{} may be misused for producing AI-generated design pieces with inaccurate interpretations of cultural symbols or artistic style. 
To reduce the misused cases, future tools could integrate features that hinder direct usage of generated images as culture-related design outcome, such as adding apparent watermarks, \wu{altering users to potential errors, or restricting the ability to download the images}. }

\penguin{\textbf{Third, the tools should ensure that users have critical thoughts or at least some inputs in each stage of the human-AI cultural co-creation process.}
\wu{Because of its powerful ability to generate high-quality outputs with a low cost, the involvement of GenAI can lead to reduced cognitive engagement, overreliance \cite{george2024erosion}, and design fixation \cite{wadinambiarachchi2024effects}. Ideation tools can mitigate these risks by encouraging users to engage in independent thinking and form their own ideas before accessing LLM assistance \cite{qin2025timing}.}
\name{} supports users in inputting their thoughts in each step of the Chinese-style visual design ideation process. 
For example, users can select suggested cultural symbols and search related Chinese paintings (DG1), analyze the paintings from multiple dimensions (DG2), and modify keywords and prompts to get generated images (DG3). 
It encourages users to spend sufficient effort in exploring the design space and coming up with ideas, instead of relying on end-to-end automation from an initial design theme to final visual designs. 
In other words, the true value of a creativity support tool lies not in relying solely on the generative models, but in empowering creators to actively engage in key stages of the creative process \cite{xiao2024typedance}. 
For more broader human-AI cultural co-creation tasks, \eg writing an essay about Chinese dragon or composing a song about Norse legends, 
\wu{we should encourage users' inputs with functions that enhance critical thinking, \eg integrating Socratic questioning techniques \cite{amquestioner_cscw25}
and providing confidence of the AI-generated content \cite{ma2023should}.
Besides, documenting and reporting the human-AI co-creation process (\eg picturing the human and AI inputs in a timeline \cite{chengbo_chart_future_chi25}) can also help users be aware of the role each party plays.}
These approaches would improve the quality of cultural designs and prove that the creators maintain ownership. 
}
\subsection{Limitations and Future Work}
First, the small sample size and the time constraint in the user study could limit the diversity and creativity of ideation outcome. 
To enhance the system’s adaptability, 
\penguin{
future work should assess our open-sourced \name{} \footnote{\url{https://github.com/ShionMing/InkIdeator}} with a more diverse group of users in their visual design tasks, 
focusing on how it affects their design outcomes and lasting impact (\eg audiences' feedback on the design). 
}
Second, \pzh{we used a baseline system that deducts the unique features of \name{}
\shiwei{with common tools that designers typically use,}
but designers may not be able to access a structured dataset of Chinese paintings and get used to GenAIs. 
Some designers would search for Chinese paintings on online platforms like RedNote, and some may gain inspiration from the representative examples in textbooks. 
Comparing \name{} to a baseline that simulates the common practices of Chinese-style visual design ideation would provide more empirical findings on its effectiveness. 
}
Third, \name{} does not support users in manually sketching their ideas in its interface. 
Future work can embed a canvas into \name{} to enable users to illustrate their thoughts via a mouse or a smart pen, which could also help iterate the design ideas.
Fourth, for the generated poems, the LLM was prompted to either return an existing poem or generate one based on the image's visual features, rather than retrieving poems from external sources.
However, the AI may lead to hallucination, which could lead to lower-quality generated poetry. 
Future work could reference the method proposed in \cite{feng2022ipoet}, which extracts features from images and maps them to a poetry database to generate more contextually grounded verses. 

\section{Conclusion}
\pzh{
In this work, we first conducted a formative study with 10 participants to understand what they learn from Chinese paintings for Chinese-style visual designs and their challenges in the ideation process. 
Then, we computationally mined the design space of 16,315 Chinese paintings regarding the cultural symbols, emotions, compositions, and styles.  
Next, we developed \name{}, an ideation support tool for Chinese-style visual design powered by Chinese paintings and generative AIs. 
\penguin{Our within-subjects study with 12 participants} demonstrated \name{}'s effectiveness in providing creative support to boost effective exploration of Chinese paintings for forming the Chinese-style visual design ideas. 
\penguin{We also had two experienced painters use \name{} for ideating Chinese painting, demonstrating its potentials for supporting other culture-related creative tasks.}
Our work provides implications for supporting culture-related visual design ideations with cultural artworks and generative AIs.
}
\pzh{
\section{
\minor{Acknowledgments}}
We used AI (in particular large language models and text-to-image generation models) for the following: analyzing the design dimensions in Chinese paintings, 
implementing \name{}'s features of recommending associated symbols and generating images given users' text input, and translating in the presented user interfaces from Chinese to English. 
Details can be found in the relevant sections. 
Authors take responsibility for the output and use of AI in this paper.
}

\minor{This work is supported in part by the General Projects Fund of the Natural Science Foundation of Guangdong Province in China with Grant No. 2024A1515012226, the Guangxi Science and Technology Program with Grant No. AB25069470, the Guangzhou Basic and Applied Basic Research Foundation with Grant No. 2024A04J6462 and the EdUHK-HKUST Joint Centre for Artificial Intelligence (JC\_AI) research scheme grant with Grant No. FB454. We thank our study participants and all experts for their support, and
all reviewers for their comments and suggestions that helped polish this paper.}
\bibliographystyle{ACM-Reference-Format}

\appendix
    \begin{figure*}[h]
    \centering
    \includegraphics[width=1.99\columnwidth]{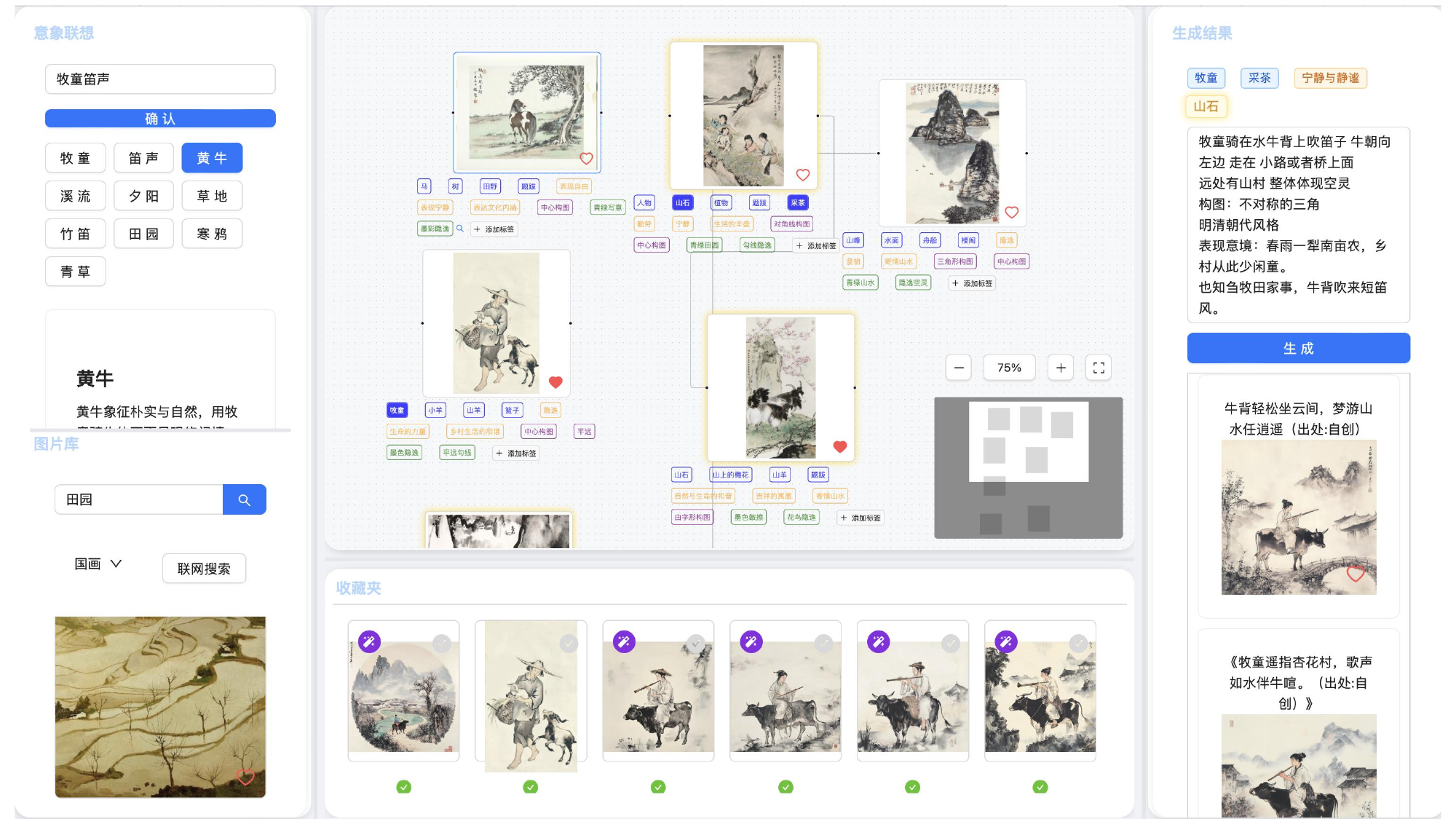}
    \caption{The screenshot of the Chinese version of \textit{InkIdeator} used for user study.}
    \label{fig:chinese}
\end{figure*}
\begin{figure*}[h]
    \centering
    \includegraphics[width=1.99\columnwidth]{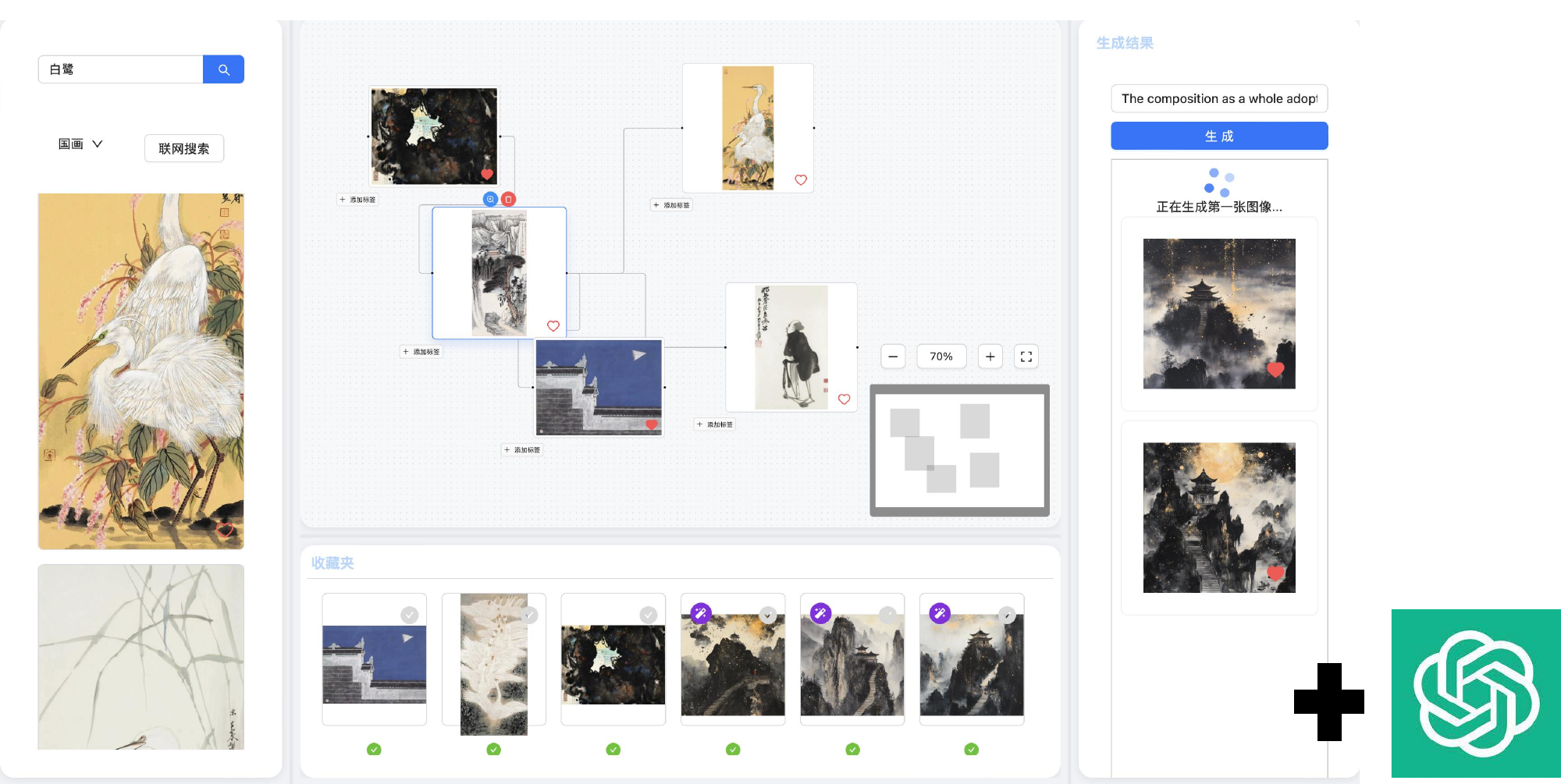}
    \caption{The screenshot of the baseline system used for user study.}
    \label{fig:baseline}
\end{figure*}

\wu{
\section{Selected bloggers} \label{sec: blogger}
\url{https://www.xiaohongshu.com/user/profile/5d05ea780000000012012a96}\\ 
\url{https://www.xiaohongshu.com/user/profile/5d215d8b0000000012037c88}\\ 
\url{https://www.xiaohongshu.com/user/profile/5a15acf711be1042a2ebee13}\\ 
\url{https://www.xiaohongshu.com/user/profile/5e686ba10000000001007092}\\ 
\url{https://www.xiaohongshu.com/user/profile/661560170000000007007ec0}\\ 
\url{https://www.xiaohongshu.com/user/profile/5f70a838000000000100a87c}\\ 
\url{https://www.xiaohongshu.com/user/profile/65293a03000000002a02a5da}\\ 
\url{https://www.xiaohongshu.com/user/profile/62503d85000000002102012b}\\ 
\url{https://www.xiaohongshu.com/user/profile/5a509abd11be1056043d8d35}\\ 
\url{https://www.xiaohongshu.com/user/profile/634672750000000018029fd5}\\ 
\url{https://www.xiaohongshu.com/user/profile/60fbe84b0000000001017b0f}\\ 
\url{https://www.xiaohongshu.com/user/profile/6139b4020000000002025dbb}\\ 
\url{https://www.xiaohongshu.com/user/profile/63ebc46b000000002702a98e}\\ 
\url{https://www.xiaohongshu.com/user/profile/63491f91000000001802e489}\\ 
\url{https://www.xiaohongshu.com/user/profile/62c570ba0000000002002fa9}\\ 
\url{https://www.xiaohongshu.com/user/profile/670e61bb000000001d020288}\\ 
\url{https://www.xiaohongshu.com/user/profile/658cd51d000000001f039f1e}. Note: This user's profile used to mostly feature traditional Chinese paintings, but they have deleted some of them recently.\\ 
}
\wu{
\section{AI-generated images evaluation procedure.}
\label{sec: evaluationappdix}
\penguin{
To evaluate the generated cultural symbols, we first prompted GPT-4o to generate 20 design themes (\eg Dunhuang Flying Apsaras, Peach Blossom Paradise), which were manually checked by the first author to ensure meaningful.  
Then, for each design theme, we prompted GPT-4o to generate ten related cultural symbols.
The inter-rater agreements (Cronbach's $\alpha$) are 0.215 and -0.142, respectively, which are relatively low but could be explained by the subjective nature of the rating task, \eg some symbols may be perceived as evocative by one rater but may not by another. 
}

\penguin{
For image generation, we prepared 20 sets of ``tags, a design intention, the first generated image and its attached poem, and \wu{two} other generated images''. 
For each set, the tags are randomly selected from our mined dataset (\autoref{tab:designspace}). Specifically, for the three dimensions (\ie cultural symbols, emotion, and style) that have larger numbers of concepts, we randomly select two tags for each dimension. 
The remaining three dimensions (\ie color tone, brushstroke, composition) each had a 0.5 chance of being selected, and if selected, one tag was randomly chosen.  
Given the tags, the remaining content of each set is generated subsequently as described in \autoref{sec:qualitative_validation}. 
To simulate the case that users may select an image prompt (\raisebox{-.18\height}{\includegraphics[width=0.86cm]{sections/fig/icon_step/step9.png}} in \autoref{fig:UserScenario}), for each of five randomly chosen sets, when generating images, the first author selected a Chinese painting, which contains at least one of the input tags, as the image prompt. 
\wu{To evaluate the impact of our craft prompts on generated image, we directly prompt Midjourney with the selected tags and additionally include the tag ``Chinese painting'' to generate images as a baseline condition.}
The inter-rater agreements for evaluation dimensions are as follows: \textit{relevance} of design intention ($\alpha = 0.028$), \textit{relevance} of generated images (\name{}: $\alpha = 0.029$, baseline: $\alpha = 0.469$), \textit{preference} (\name{}: $\alpha = 0.340$, baseline:  $\alpha = 0.225$), \textit{aesthetics} (\name{}: $\alpha = 0.610$, baseline: $\alpha = 0.331$), \textit{diversity} (\name{}: $\alpha = 0.407$, baseline: $\alpha = 0.742$), \textit{relevance} of poem ($\alpha = 0.529$).
We accounted the low inter-rater agreements on some items (\eg the relevance of intention, diversity of images) for the subjective nature of the rating tasks, as the raters 
evaluate it based on their understandings. \wu{Raters also have different standards, like S11's score relatively lower compared to the other two raters.}
}
}
\end{document}